\documentclass[preprint,11pt,authoryear]{elsarticle}
\usepackage[margin=2.5cm]{geometry}
\pdfoutput=1
\usepackage{amsmath}
\usepackage{amsthm}
\usepackage{float}
\usepackage{tabularx,booktabs}
\newtheorem{theorem}{Theorem}
\newtheorem{lemma}{Lemma}
\newtheorem{definition}{Definition}
\usepackage{float}
\usepackage[ruled,vlined]{algorithm2e}
\usepackage{algpseudocode}
\usepackage{caption}
\usepackage[x11names]{xcolor}
\makeatletter
\DeclareRobustCommand{\change}{%
  \@bsphack
  \leavevmode
  \color{red}%
  \@esphack
}
\DeclareRobustCommand{\stopchange}{%
  \@bsphack
  \normalcolor
  \@esphack
}
\makeatother

\usepackage{amssymb}

\journal{The Journal of Machine Learning with Applications}
\newpageafter{author}
\begin{document}

\begin{frontmatter}


\title{Portfolio Optimization based on Neural Networks Sensitivities from Assets Dynamics respect Common Drivers}
\author[label1]{Alejandro Rodriguez Dominguez}
\affiliation[label1]{
             organization={Miralta Finance Bank S.A.},
             addressline={Plaza Manuel Gomez Moreno,2, 17-A},             city={Madrid},             postcode={28020},            state={Spain},
                    country={arodriguez@miraltabank.com}}


\begin{abstract}
We present a framework for modeling asset and portfolio dynamics, incorporating this information into portfolio optimization. We define drivers for asset and portfolio dynamics and their optimal selection. For this framework, we introduce the Commonality Principle, providing a solution for the optimal selection of portfolio drivers as the common drivers. Portfolio constituent dynamics are modeled by Partial Differential Equations, and solutions approximated with neural networks. Sensitivities with respect to the common drivers are obtained via Automatic Adjoint Differentiation. Information on asset dynamics is incorporated via sensitivities into portfolio optimization. Portfolio constituents are embedded into the space of sensitivities with respect to their common drivers, and a distance matrix in this space called the Sensitivity matrix is used to solve the convex optimization for diversification. The sensitivity matrix measures the similarity of the projections of portfolio constituents on a vector space formed by common drivers’ returns and is used to optimize for diversification on both idiosyncratic and systematic risks while adding directionality and future behavior information via returns dynamics. For portfolio optimization, we perform hierarchical clustering on the sensitivity matrix. The clustering tree is used for recursive bisection to obtain the weights. To the best of the author’s knowledge, this is the first time that sensitivities’ dynamics approximated with neural networks have been used for portfolio optimization. Secondly, that hierarchical clustering on a matrix of sensitivities is used to solve the convex optimization problem and incorporate the hierarchical information of these sensitivities. Thirdly, public and listed variables can be used to obtain maximum idiosyncratic and systematic diversification by means of the sensitivity space with respect to optimal portfolio drivers. We reach over-performance in many experiments with respect to all other out-of-sample methods for different markets and real datasets. We also include a recipe for the methodology to increase performance even further, and tackle the main issues in portfolio management such as regimes, non-stationarity, overfitting, and selection bias.
\end{abstract}

\end{frontmatter}


\section{Introduction}
\label{sec:sample1}
Portfolio Management relies on diversification to reduce portfolio risk since \citep{10.2307/2975974} Modern Portfolio Theory. Sharpe and Lintner introduced the Capital Asset Pricing Model (CAPM) stating the limit for diversification \citep{RePEc:bla:jfinan:v:19:y:1964:i:3:p:425-442,10.2307/1924119} in systematic risk allowing only for idiosyncratic diversification. The Arbitrage Pricing Theory (APT) introduced by \citep{ross1976arbitrage, RePEc:rje:bellje:v:8:y:1977:i:spring:p:23-40} solved this drawback by adding exogenous information with common risk factors and idiosyncratic factors driving asset returns. It has developed extensive literature on factor models, risk premia, and smart beta until nowadays. Bayesian approaches introduced by \citep{Black7} added exogenous information, causality inference, and systematic diversification with probabilistic models which have been extended up to nowadays with bayesian networks as in \citep{doi:10.2469/faj.v48.n5.28, RePEc:pra:mprapa:30534}. There is already an extensive literature on asset and portfolio dynamics modeled with stochastic and partial differential equations used by practitioners in hedging, risk management, and derivative pricing but there are far fewer applications in portfolio optimization. For this reason, the main contribution of this paper is to incorporate the sensitivities of portfolio constituents with respect to their common drivers approximated with neural networks and optimize diversification with respect to hierarchical representations of their sensitivity dynamics.\par
If we model the true asset dynamics with differential equations using its drivers as exogenous variables, and not risk factors as in \citep{ 10.1016/j.eswa.2021.116308, 4744738} or functions of individual variables, and we use assets and drivers’ time series to approximate the solution of differential equations and partial derivatives with neural networks. Consequently, we can incorporate the sensitivity information into the portfolio optimization allowing us to include asset and portfolio dynamics and hence future behavior, including true risk and return dynamics approximated with public and listed data instead of statistical risk factors, for diversification. We model asset dynamics with Partial Differential Equations (PDEs), approximating their solution and sensitivities with Automatic Adjoint Differentiation (AAD) in neural networks. \par
For the inclusion of the sensitivities in portfolio optimization, portfolio constituents need to be embedded in a space of sensitivities with respect to their common drivers, which we refer to in the commonality principle as the optimal portfolio drivers in terms of causality and persistence. The geometric space formed by the sensitivities of portfolio constituents with respect to their common drivers is a new pathway to retaining the maximum amount of idiosyncratic risk from traditional mean-variance methods while adding systematic risks for portfolio diversification. Driver optimality is based on the largest correlation in terms of persistence and different lags. For each portfolio constituent, we train a neural network with the common drivers as inputs and constituents as targets, for different lags and network parameters selecting the best architecture in terms of accuracy. The same exercise is performed for other constituents with the same common drivers, and the best architecture is selected for each constituent, and sensitivities of each portfolio constituent with respect to the common drivers are obtained with AAD. These sensitivities are discrete functions of the training data, and averages or other statistics are used to resume each of the sensitivities. \par
As a result, we obtain a vector of sensitivity values' averages with respect to the common drivers for each constituent and embed them in that vector space. A distance matrix called the Sensitivity matrix is computed in this space and used for portfolio optimization. In order to solve the convex optimization for diversification in this space of projections of constituents with respect to optimal portfolio drivers we apply hierarchical clustering and take advantage of the hierarchical information of these projections, as in \citep{Prado2016BuildingDP} with correlations (cosines or constituents' projections). Hierarchical clustering is applied to a positive semi-definite neighbor of the sensitivity matrix with the single-linkage algorithm and a distance metric which is a function of the sensitivity matrix. Then, clustered items are sorted by distance and we utilize the Recursive Bijection technique from \citep{Prado2016BuildingDP} with our clustering tree and clusters’ covariances. In \citep{Prado2016BuildingDP}, recursive bijection is applied with a clustering tree from the single-linkage algorithm and correlation distance. We perform experiments on two time series datasets sourced from Bloomberg, focusing on the European and US equity markets. In many experiments, overperformance is reached with respect to the rest of the out-of-sample methods.\par
The main contributions of this work are to incorporate sensitivity dynamics information approximated with neural networks into portfolio optimization. Secondly, use hierarchical clustering on the sensitivity matrix to solve the convex optimization problem and incorporate hierarchal information from these sensitivities, naming it Hierarchical Sensitivity Parity (HSP). Thirdly, develop a new way to obtain maximum idiosyncratic and systematic diversification by means of the sensitivity space with respect to optimal portfolio drivers (common drivers). \par
The paper is organized as follows: Section 2 presents a literature review on portfolio theory and diversification, and applications utilizing machine learning, time series analysis, and predictability;  Section 3 details the model for asset and portfolio dynamics, the optimal selection of drivers as common drivers, and a geometric explanation of our method; Section 4 outlines our method and explains the HSP algorithm; In Section 5 we evaluate the overall performance of our methodology by analyzing real financial data belonging to two different markets: the US and European stocks market; Section 6 presents management insights regarding the choice of the strategy hyperparameters and experiments; Section 7 concludes the paper with the main insights and a discussion on further lines of research.\par

\section{Related Literature}
\label{sec:sample2}
\subsection{Portfolio Theory and Diversification}
 In the field of portfolio management, there has been an extensive list of portfolio selection methods since \citep{10.2307/2975974} introduced Modern Portfolio Theory. The concept of diversification starts with Markowitz, and his question: What explains diversification?. \citep{RePEc:oup:restud:v:25:y:1958:i:2:p:65-86} created the two-factor model (expected value and variance) and the mean-variance framework, that relies on investors’ rationality and risk aversion. Then, the Capital Asset Pricing Model (CAPM) is derived from the two-factor approach in \citep{RePEc:bla:jfinan:v:19:y:1964:i:3:p:425-442,10.2307/1924119} as equilibrium for the mean-variance framework starting the asset pricing literary movement. Lintner developed formal proofs for the justification of the use of variance as a risk measure, by connecting the assumptions of uncertainty and rational behavior, using expected utility theory, probability, market risk conditions and assumptions on the market competition and expectations \citep{RePEc:bla:jfinan:v:20:y:1965:i:4:p:587-615}. The next movement comes from the empirical analysis of the two-factor approach, the Efficient Capital Markets Hypothesis (EMH), and Rational Expectations. In \citep{RePEc:bla:jfinan:v:25:y:1970:i:2:p:383-417} a framework to test (EMH) is developed. Empirical evidence suggests CAPM delivers poor results for diversification. Others, as in \citep{RePEc:eee:jfinec:v:4:y:1977:i:2:p:129-176}, argue that CAPM is not testable for diversification as market portfolio is unobserved.\par
Further developments come from \citep{RePEc:rje:bellje:v:4:y:1973:i:spring:p:141-183}, with a dynamic extension of the traditional CAPM from Sharpe and Lintler. Investment opportunity evolves over time by means of the stochasticity of the risk-free rate and the Markov theorem allows for the first two moments to explain portfolio selection, both with restrictive assumptions in continuity, free-arbitrage, and others. This movement further developed in all theories behind derivatives pricing, however, there is strong empirical evidence that suggests all these models have unrealistic assumptions as in \citep{HAUG201197}.\par
\subsection{Machine Learning, Time Series analysis, Predictability}
Machine Learning models have been extensively used to detect latent factors that drive market returns. In \citep{ 10.1016/j.eswa.2021.116308,4744738} the authors focus on risk factors featuring extraction for portfolio optimization. In contrast, our work focuses on the true asset and portfolio dynamics approximation with public and listed information of drivers, instead of statistical or hidden factors and the use of sensitivity dynamics information for diversification, adding directionality and the future behavior of assets to risk diversification. Dimensionality reduction, clustering, and regression techniques are applied to compute distance matrices that can outperform in a convex portfolio optimization setting. References include \citep{marti2016clustering} with a survey about clustering financial time series, \citep{Prado2016BuildingDP} Hierarchical Risk Parity (HRP) method, in which hierarchical structure of the correlation matrix is used to improve portfolio selection. In \citep{avellaneda2020hierarchical}, authors apply hierarchical Principal Component Analysis (PCA) and spectral properties of the correlation matrix to portfolio construction. On the other hand, many developments have been carried out in regime switch detection and change point detection and the benefits for Portfolio Management. In \citep{RePEc:eee:econom:v:143:y:2008:i:2:p:263-273} authors introduce regime-switching models, in \citep{doi:10.1080/14697688.2017.1342857} authors construct a regime-dependent portfolio and show that forecasting regimes and incorporating this into the process can improve risk-adjusted returns. \par
In \citep{RePEc:eee:phsmap:v:515:y:2019:i:c:p:671-681} authors introduce network topology to portfolio optimization, \citep{satone2021fund2vec} develop Fund2Vec, a new way of using embeddings for building portfolios of mutual funds. In \citep{RePEc:eee:phsmap:v:335:y:2004:i:3:p:629-643,RePEc:sfi:sfiwpa:500053,2002,pafka2004exponential} authors focus on Random Matrix Theory (RMT) for portfolio optimization. RMT is used to improve results by denoising and detuning the correlation matrix. Bayesian inference for portfolio management was introduced in \citep{doi:10.2469/faj.v48.n5.28}. Since then, work has been done by \citep{Bevan98usingthe,RePEc:pal:assmgt:v:1:y:2000:i:2:d:10.1057-palgrave.jam.2240011}, and \citep{citekey} with a Black-Litterman model extension. Bayesian Networks have added structure to the Bayesian inference paradigm as in \citep{RePEc:pra:mprapa:30534}. These techniques incorporate causality into portfolio selection. \par

Neural networks have been used for portfolio optimization in Minimax and bi-objective portfolio selection based on collaborative neurodynamic optimization by \citep{8948344}, and decentralized robust portfolio optimization based on cooperative-competitive multiagent systems by \citep{9484812}. Our approach uses a neural network to approximate asset and portfolio dynamics which can be modeled by PDEs or Stochastic Differential Equations (SDEs) with drivers as exogenous variables, and partial derivatives are approximated with sensitivities from neural networks with Automated Adjoint Differentiation (AAD). Neural networks are used as feature extractors of their dynamics and not as solvers of the convex optimization problem. The method we use for obtaining the sensitivities is focused on Derivative-based local methods, particularly on Adjoint modeling and Automated Differentiation (AAD). AAD is a common method for extracting derivatives and system sensitivities. In financial applications, \citep{huge2020differential} introduced Differential Machine Learning for derivative pricing and hedging. Additionally, references for sensitivity analysis with neural networks in other sciences include \citep{zesmi1999AnAN,doi:10.1177/089443939100900304,OLDEN2004389,Scardi1999DevelopingAE}, with a survey to be found in \citep{pizarroso2021neuralsens}, R package for time series.\par

\section{Theoretical Framework}
\subsection{A Model for Asset and Portfolio Dynamics}

Our method focuses on asset dynamics that can be modeled by unknown PDEs with independent variables as drivers. The PDEs are approximated with neural networks with time series data and sensitivities of assets with respect to the drivers which are obtained from these neural networks. \par
If we model financial assets as first-order PDEs with y being the asset and x a vector of drivers causing its dynamics, this PDE is unknown and analytically unsolvable, and its solution could be represented as:

\begin{equation}
y\left(t\right)=F(\frac{\partial y\left(t\right)}{{\partial x}_1\left(t\right)},\frac{\partial y\left(t\right)}{{\partial x}_2\left(t\right)},\dots,\frac{\partial y\left(t\right)}{{\partial x}_N\left(t\right)},\frac{{\partial x}_1}{\partial t},\frac{{\partial x}_N}{\partial t},\frac{\partial y}{\partial t},x_1,\ \dots,x_N)    
\label{PDE1}
\end{equation}

F can be non-linear, but in the case of being linear it can be expressed as:

\begin{equation}
y\left(t\right)=x_1\left(t\right)\frac{\partial y\left(t\right)}{{\partial x}_1\left(t\right)}+x_2\left(t\right)\frac{\partial y\left(t\right)}{{\partial x}_2\left(t\right)}+\dots+x_N\left(t\right)\frac{\partial y\left(t\right)}{{\partial x}_N\left(t\right)}+\frac{\partial y}{\partial t}+\frac{{\partial x}_1}{\partial t}+\dots +\frac{{\partial x}_N}{\partial t}
\label{PDE2}
\end{equation}

We can approximate a PDE solution of the type in (\ref{PDE1}) with time series data and a neural network because they are universal approximators, as demonstrated in \citep{Cybenko1989ApproximationBS}. Below is shown an example of the functional approximator of a dynamical system like (\ref{PDE2}) and its components (Figure 1 for illustration), following \citep{AntonioMunozSanRoque}, adapted to our problem:
\begin{equation}
d\left[k\right]=g\left(d^{\left\{k-1\right\}},u^{\left\{k\right\}},\varepsilon^{\left\{k-1\right\}}\right)+\varepsilon\left[k\right]
\end{equation}

Where:\par
\begin{itemize}
	\item g is a non-linear function 
	\item $d[k]$ is the prediction of asset return at time k
	\item $d^{\left\{k-1\right\}}=\ \left[d\left[k-1\right],\ d\left[k-2\right],\dots\right]^T$is a vector containing the asset return in times k-1 and backwards.
	\item $u^{\left\{k\right\}}=\ \left[u\left[k-1\right],\ u\left[k-2\right],\dots\right]^T$ is a vector containing the exogenous entries at times k and backward (the asset drivers).
	\item $\varepsilon^{\left\{k\right\}}=\ \left[\varepsilon\left[k-1\right],\ \varepsilon\left[k-2\right],\dots\right]^T$ is a vector with the White-Noise(WN) realizations at the asset prediction at time k-1 and backwards.
	\item $\varepsilon\left[k\right]$ is the WN realization at the asset prediction at time k. 

\end{itemize}

The system that represents the PDE solution in (\ref{PDE2}) for our case, can be approximated with a neural network $F(x,w)$ in which inputs $x(k)$ are, $u^{\left\{k\right\}}$ (feed-forward networks), or $u^{\left\{k\right\}}$, $d^{\left\{k-1\right\}}$ and $\varepsilon^{\left\{k\right\}}$ (recurrent networks), so that different configurations of neural networks can be used to approximate different modeling PDEs. Here the focus is specifically on feed-forward networks, with inputs, $u^{\left\{k\right\}}$, the asset drivers, and output, $d[k]=y(k)$, the prediction of asset returns, as in Figure 1. Once true dynamics are approximated with exogenous drivers, the sensitivities of the assets with respect to their drivers are obtained via AAD.

\begin{figure}
\label{figure1}
	\centering
	\includegraphics[width=65mm]{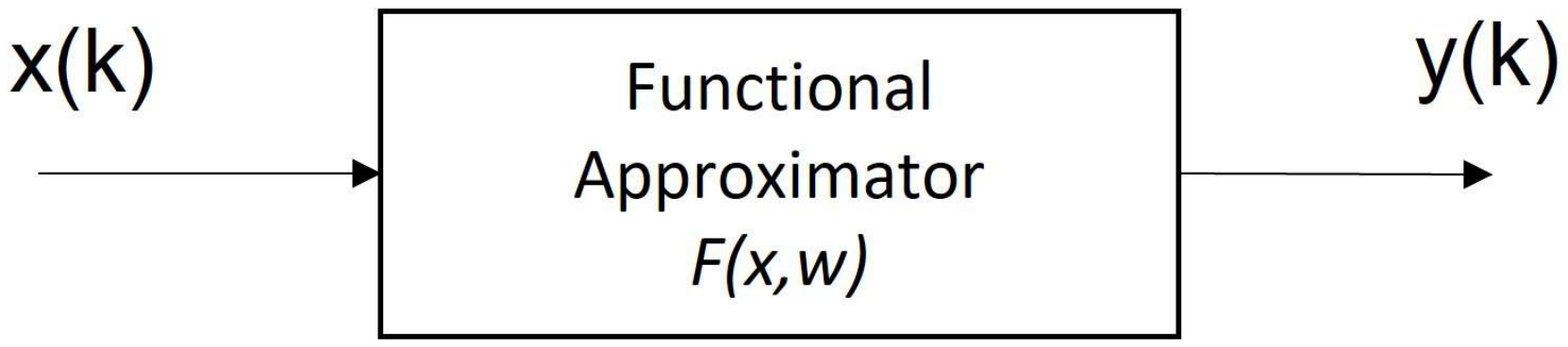}
	\caption{Functional Approximator Scheme}
\end{figure}

\subsection{Optimal drivers selection}
We select drivers with two goals, to maximize the accuracy when assets and portfolio dynamics are approximated, and to utilize sensitivity information to project portfolio constituents into a sensitivity space, from which we can diversify with respect to this dynamic risk and return information. If we define drivers as exogenous market or economic variables daily quoted as public information which is the cause of asset dynamics, we can define their optimal selection for assets and portfolios. 
\begin{definition} Drivers Optimality. A driver is optimal for an asset if it is:
\begin{itemize}
\item Optimal in persistence, the amount of time it remains as a driver.
\item Optimal in the probability of causality. As causality cannot be guaranteed, we speak in terms of probabilities and an optimal driver must maximize the probability of causing the asset dynamics.
\end{itemize}
\end{definition}
\begin{definition}
Specific Drivers are the optimal drivers for individual assets (portfolio constituents).
\end{definition}
Optimal drivers’ selection allows us to better approximate asset dynamics and sensitivities. We demonstrate here that we must use the set of most common specific drivers among all constituents to use constituents’ sensitivity information in portfolio optimization so that the space of sensitivities in which constituents are projected allows for optimal diversification. Furthermore, this set of common drivers is also the set of optimal portfolio drivers based on the definition of drivers’ optimality. \par
The following theorem and proof seem too obvious to highlight, but it is necessary for our methodology both for better portfolio dynamics approximation and to achieve greater diversification. The theorem may not be a contribution, but is a necessary link between different fields, to the best of the author’s knowledge the first time it has been mentioned in the literature for use in our novel methodology. 
\begin{theorem}
The Commonality Principle for Portfolio Drivers states that optimal drivers for a portfolio are the specific drivers that are repeatedly selected for the greatest number of portfolio constituents, both in terms of persistence and probability of causality.
\end{theorem}
\begin{proof}
For the Commonality Principle, we first need to prove that the common drivers, the most repeatedly selected as specific drivers among all constituents, are the most persistent so that they are optimal for a portfolio. For this, we can make use of Modern Portfolio Theory (MPT).\par
Drivers contribute to the risks of an asset by providing the source of its dynamics. We can categorize risks into idiosyncratic and systematic, and specific drivers can contribute to both types of risks. Most specific drivers from a particular portfolio constituent will contribute to its own idiosyncratic risk, which does not affect other constituents. But also, for a particular constituent, some specific drivers may contribute to the systematic risk of this individual asset and of other constituents too. As in MPT, it can be stated that the amount of systematic risk explained by all specific drivers from all constituents is not maximal, because there is a significant number of specific drivers focused on idiosyncratic risks alone. Or equivalently, we are not restricting the problem by looking only for drivers of systematic risks.\par
But, when looking for the portfolio drivers that follow the commonality principle, we are restricting the focus to the specific drivers which are most common. In this case, we do maximize the amount of systematic risk from each and all portfolio constituents, which is explained by any possible subset of drivers. This systematic risk is the most common among all portfolio constituents and is the most persistent for the portfolio, as seen in MPT. Therefore, optimality in driver persistence for a portfolio is guaranteed with common drivers’ selection, which contributes the maximum amount of systematic risk to a particular portfolio.
\end{proof}

\begin{proof}
We secondly need to provide proof that the common drivers are the set among the specific drivers for all constituents that have the highest probability of causality of portfolio dynamics. We do this by demonstrating that the maximum probability of causality for a portfolio, given any possible selection of drivers, is obtained in the optimal choice of drivers following the Commonality Principle. There is no guarantee of causality, so we must use the probability of causality. \par

\begin{itemize}
\item Probability of causality for an asset or a portfolio, given a set of drivers, is defined as the probability that their dynamics are caused by this set of drivers.

\item	For each constituent $1,\dots, N$ of a portfolio and its set of specific drivers (M specific drivers SD1 for Asset 1$,\dots,$ W specific drivers SDN for Asset N) there is a probability of causality, probability of causing the dynamics at p time steps in the future given by the following vector X, using Judea Pearl notation \citep{neuberg_2003}:

\begin{equation}
\begin{split}
P\left({Asset1}_{t+p}\middle|do(\left[SD1_1,\dots SD1_M\right]_t\right))\le X_1,\dots,\\ P\left({AssetN}_{t+p}\middle|do(\left[SDN_1,\dots SDN_W\right]_t\right))\le X_N    
\end{split}
\end{equation}

 \item	At a portfolio level we have:
\begin{equation}P\left({Portfolio}_{t+p}\middle|\left[CD_1,\dots CD_M\right]_t\right)\le Y\end{equation}   with M common drivers CD.  
  \item	To prove the principle (drivers’ optimality) for the probability of causality we need to verify the following proposition is almost surely true. We will show that it is, but only in the special case that we focus on a portfolio level, which is coincidentally our only interest in portfolio optimization. $\forall \ Portfolio=[Asset1,\dots,AssetN]$ and specific drivers $SD=[SD1_1,\dots SD1_M,\dots,SDN_1,\dots SDN_W]$ and common drivers $CD=[CD_1,\dots CD_M]$:

\begin{equation}
\begin{split}
P\left({Portfolio}_{t+p}\middle|\left[CD_1,\dots CD_M\right]_t\right)=\\
P\left({Asset1}_{t+p}\cap\dots\cap{AssetN}_{t+p}\middle|\left[CD_1,\dots CD_M\right]_t\right)>\\
P\left({Asset1}_{t+p}\middle|\left[SD1_1,\dots SD1_M\right]_t\right)*\dots*P\left({AssetN}_{t+p}\middle|\left[SDN_1,\dots SDN_W\right]_t\right)=\\P({Asset1}_{t+p})*\dots*P({AssetN}_{t+p})
\end{split}
\end{equation}

\item For the proof we make use of \citep{Reichenbach1956-REITDO-2} concept of the Common Cause Principle (CCP). Suppose that events A and B are positively probabilistically correlated: \begin{equation} p(A \cap B) > p(A)p(B) 
\label{equation9}
\end{equation} Reichenbach’s Common Cause Principle states that when such a probabilistic correlation between A and B exists, this is because one of the following causal relations exists: A is a cause of B; B is a cause of A; or A and B are both caused by a third factor, C. In the last case, the common cause C occurs prior to A and B, and must satisfy the following four independent conditions:
\begin{equation} p(A \cap B|C) = p(A|C)p(B|C)
\label{equation10}
\end{equation}
\begin{equation} p(A \cap B|\overline{\rm C}) = p(A|\overline{\rm C})p(B|\overline{\rm C}) 
\label{equation11}
\end{equation}
\begin{equation} p(A |C) > p(A |\overline{\rm C}))
\label{equation12}\end{equation}
\begin{equation} p(B|C) > p(B|\overline{\rm C})
\label{equation13}\end{equation}  
$\overline{\rm C}$ denotes the absence of event C (the negation of the proposition that C happens) and it is assumed that neither C nor $\overline{\rm C}$  has probability zero. Condition (\ref{equation10}) states that A and B are conditionally independent, given C. In Reichenbach’s terminology, C screens A off from B. Condition (\ref{equation11}) states that $\overline{\rm C}$  also screens A off from B. Conditions (\ref{equation12}) and (\ref{equation13}) state that A and B are more probable, conditional on C, than conditional on the absence of C. These inequalities are natural consequences of C being a cause of A and of B. Together, conditions (\ref{equation10}) through (\ref{equation13}) mathematically entail (\ref{equation9}). The common cause can thus be understood to explain the correlation in (\ref{equation9}) \citep{Reichenbach1956-REITDO-2}.

\item For the general (CCP) case that the correlated effects are random variables like ours. Suppose X and Y are random variables that are correlated, ie, there exist  $x_{i}$ and $y_{j}$ such that \begin{equation} p(X = x_{i} \cap Y = y_{j} ) \neq p(X = x_{i}) p(Y = y_{j})\end{equation} 
Then there exists a set of variables $Z_{1},\dots,Z_{M}$ so that each variable is the cause of X and Y, and 
\begin{equation}
\begin{split}
p(X = x_{i} \cap Y = y_{j}| Z_{1} = z_{k_{1}},\dots, Z_{m} = z_{k_{m}}) =\\
p(X = x_{i}| Z_{1} = z_{k_{1}},\dots, Z_{m} = z_{k_{m}})p(Y = y_{j}| Z_{1} = z_{k_{1}},\dots, Z_{m} = z_{k_{m}}) \end{split}
\end{equation}

We check how the independent conditions are met for our case (CD=Common Drivers):

\begin{equation} p(Asset_X \cap Asset_Y|CD) = p(Asset_X|CD)p(Asset_Y|CD)
\label{equation20}
\end{equation}
\begin{equation} p(Asset_X \cap Asset_Y|\overline{\rm CD}) = p(Asset_X|\overline{\rm CD})p(Asset_Y|\overline{\rm CD}) 
\label{equation21}
\end{equation}
\begin{equation} p(Asset_X |CD) > p(Asset_X |\overline{\rm CD}))
\label{equation22}
\end{equation}
\begin{equation} p(Asset_Y|CD) > p(Asset_Y|\overline{\rm CD})
\label{equation23}
\end{equation}

This mathematically entails:

\begin{equation} p(Asset_X \cap Asset_Y|CD) > p(Asset_X)p(Asset_Y)
\label{Equation24}
\end{equation}

The common drivers (common cause) can thus be understood to explain the correlation between assets in the portfolio. The common cause must occur prior, which is our case for the common drivers and asset dynamics. The generalization of CCP is given by the Causal Markov Condition (CMC): A variable X is independent of every other variable (except X’s effects) and conditional on all its direct causes. CMC can be applied to all pairs of portfolio constituents as a generalization of CCP given the subset of common drivers such that (\ref{equation20}-\ref{equation23}) holds. For that, we need the common driver’s subset to be the direct cause of portfolio constituent dynamics, which we approximate with correlation, by making use of the CCP for the particular case that the common cause (common drivers) is, at most, the same for all portfolio constituents.

\begin{equation}
\forall X\in A, \forall S_X\in CD, P(X|do(S_X)) > P(X|do(\sim S_X))
\label{Equation25}
\end{equation}
\begin{equation}
\begin{split}
 \forall\ X,Y\in A,\ \forall\ S_X,S_Y\in CD\ |\ \\ 
 \left[P(X\ |\ do(S_X))\ >\ P(X\ |\ do(\sim S_X))\right] \land\ \left[P(Y\ |\ do(S_Y))\ >\ P(Y\ |\ do\left(\sim S_Y\right))\right]\\
 \rightarrow\ \left[P\left(X\cap Y\middle| S_X,S_Y\right)=
 P\left(X\middle| S_X\right)P\left(Y\middle| S_Y\right)\right]=
 \left[P\left(X\cap Y\middle| S\right)= P(X)P(Y)\right],\\ S \equiv S_X\equiv S_Y
 \label{Equation26}
\end{split}
\end{equation}

\begin{equation}
\begin{split}
\left\{\forall X\subseteq A,\forall\ S\subseteq C D
\left[P(X|do(S))>P(X|do(\sim S))\right]\right\}\\
\longleftrightarrow\\
[P\left(X_1\cap X_2,\cap \dots \middle| S_{X_1},S_{X_2}, \dots\right)=
P\left(X_1\middle| S_{X_1}\right)P\left(X_2\middle| S_{X_2}\right), \dots]\equiv\\
[P\left(X_1\cap X_2,\cap \dots \middle|S \right)=P{(X}_1)P{(X}_2)\dots]\\
\forall X_1,\ X_2, \dots \in X,\forall S_{X_1}\equiv S_{X_2}\equiv,\dots \equiv S
\label{Equation27}
\end{split}
\end{equation}

\item In (\ref{Equation25}), we show that for any asset X, there exists a set of common drivers that cause its dynamics optimally in probability, using Judea Pearl notation \citep{neuberg_2003}. In (\ref{Equation26}), we show for a portfolio of two assets, and its common drivers’ selection, how, if they have the highest probability of causality for the assets’ dynamics, and both $S_X$ and $S_Y$ are equivalent from the commonality principle, this entails that CCP conditions and (\ref{Equation24}) are met. This means that the common drivers are the common source of causality of portfolio constituents’ dynamics, they are the greatest source in the probability of causality for portfolio dynamics, and they explain the correlation between portfolio constituents, by applying \citep{Reichenbach1956-REITDO-2}. 
\item In (\ref{Equation27}), we show the generalization for any combination of assets (portfolio). Here, the implication goes two ways in that, for any portfolio of assets, their common drivers being the source of the highest probability of causality of portfolio dynamics (not their constituents), imply CCP conditions and (\ref{Equation24}) are met. But, if CCP and (\ref{Equation24}) are met, which occurs only in the case that the common drivers are selected based on the commonality principle, which in turn makes the equivalence in sets S possible, CCP conditions and (\ref{Equation24}) imply that they are the greatest source in the probability of causality for portfolio dynamics. This is true for any combinations of assets (X), and common drivers set (CD) chosen as in the commonality principle. But also, like in (\ref{Equation26}), this means that common drivers explain the correlation between portfolio constituents by applying \citep{Reichenbach1956-REITDO-2}, which justifies our selection of common drivers (common cause) as the drivers that are most correlated to the greatest number of portfolio constituents.
\end{itemize}
\end{proof}

\subsection{Diversification and Geometry}
\label{sectionGeometry}
\begin{figure}
    \label{figure2}
	\centering
	\includegraphics[width=65mm]{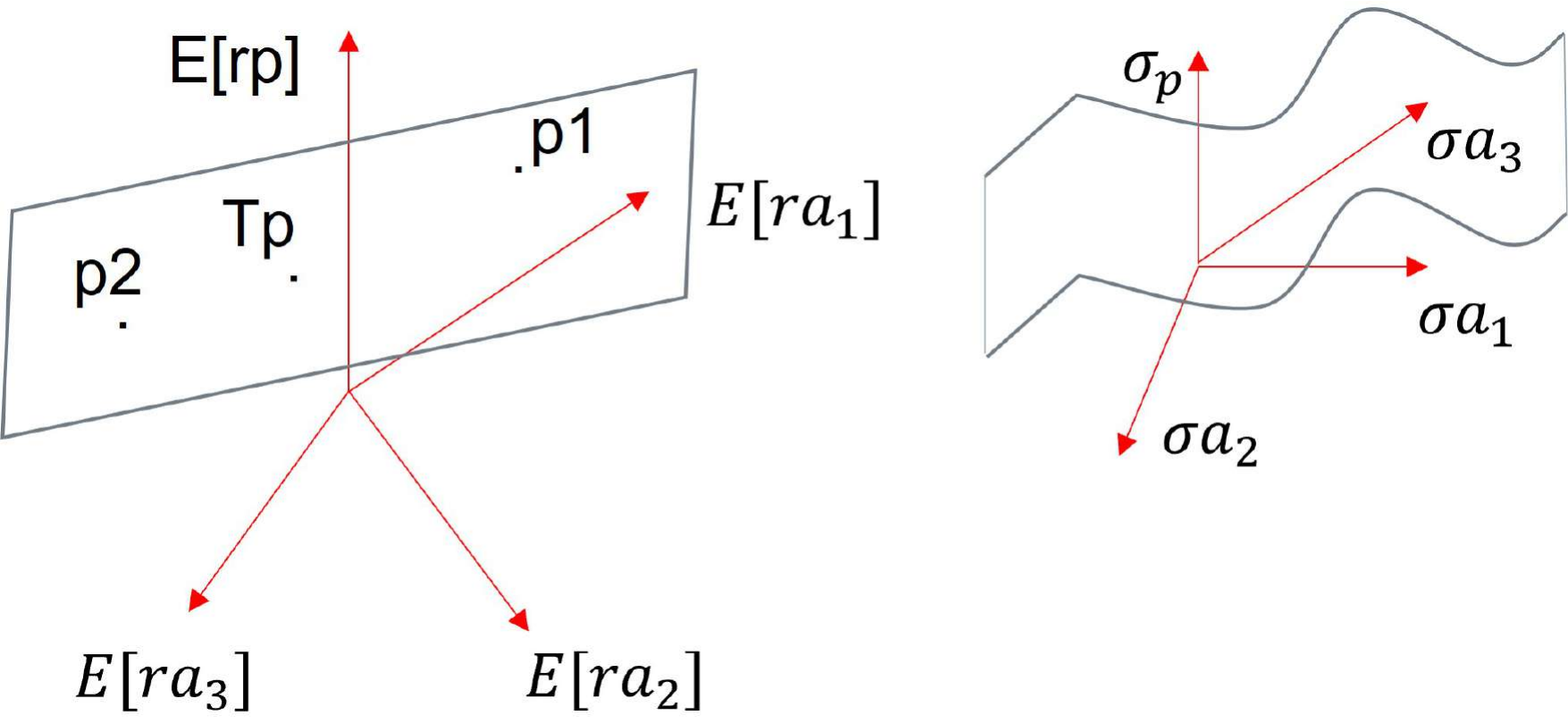}
	\caption{Mean-variance portfolio expected return hyperplane and risk hypersurface}
	\label{figure3}
	\includegraphics[width=75mm]{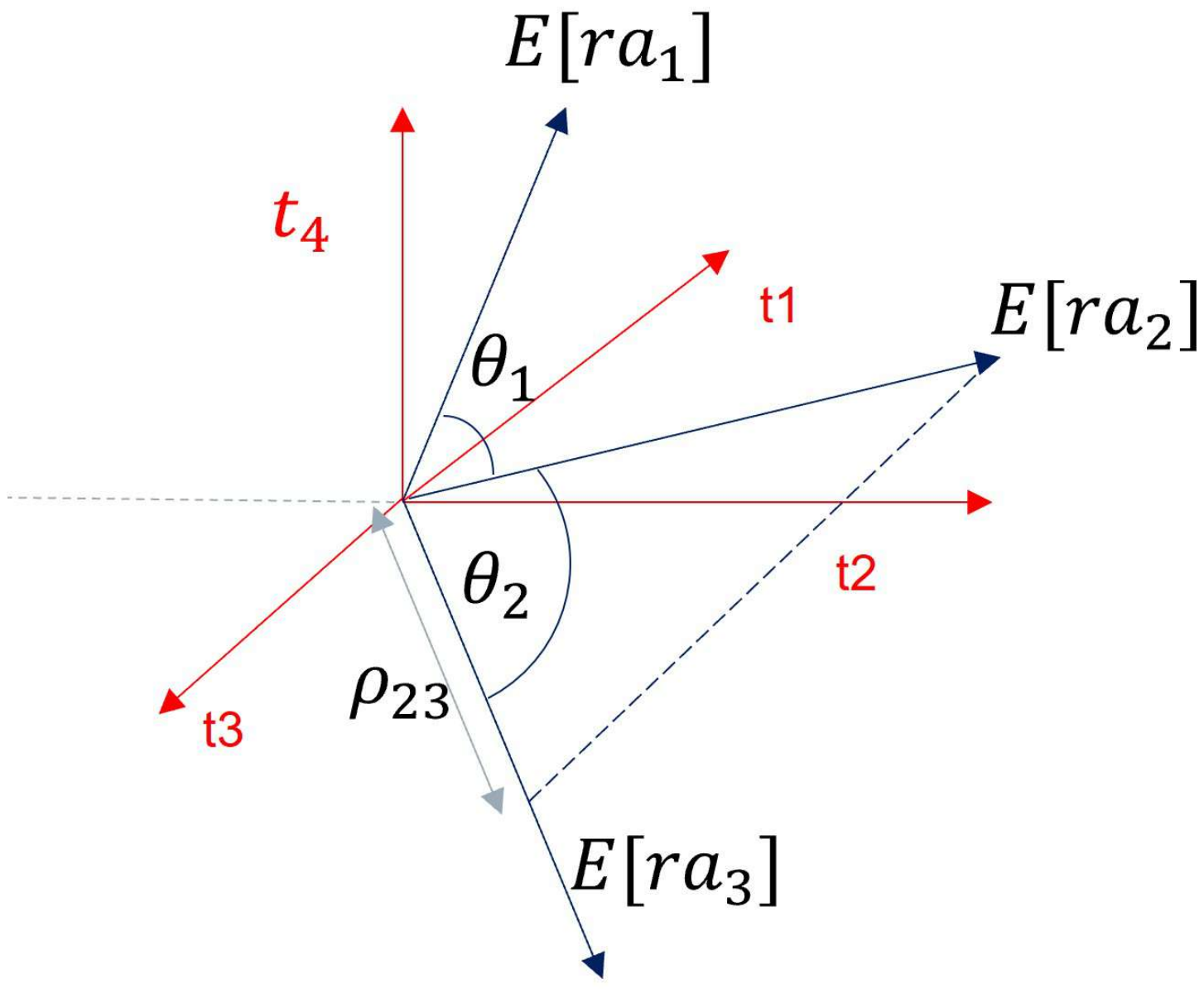}
	\caption{Mean-variance setup represented in the time-dimensional space}
\end{figure}
\begin{figure}[ht]
    \label{figure4}
	\centering
	\includegraphics[width=65mm]{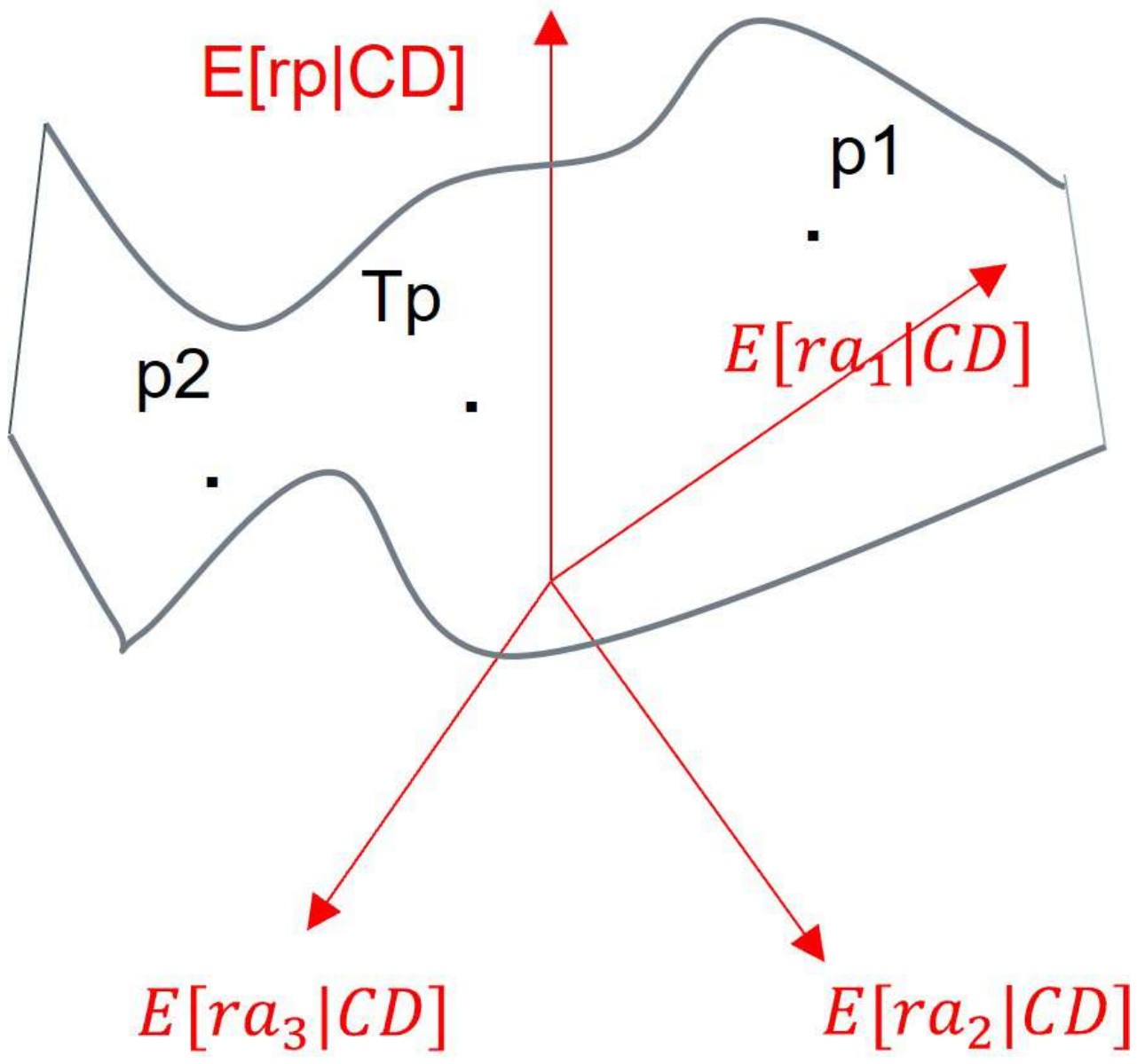}
	\caption{Conditional portfolio expected returns hypersurface}
	\label{figureCM}
	\includegraphics[width=130mm]{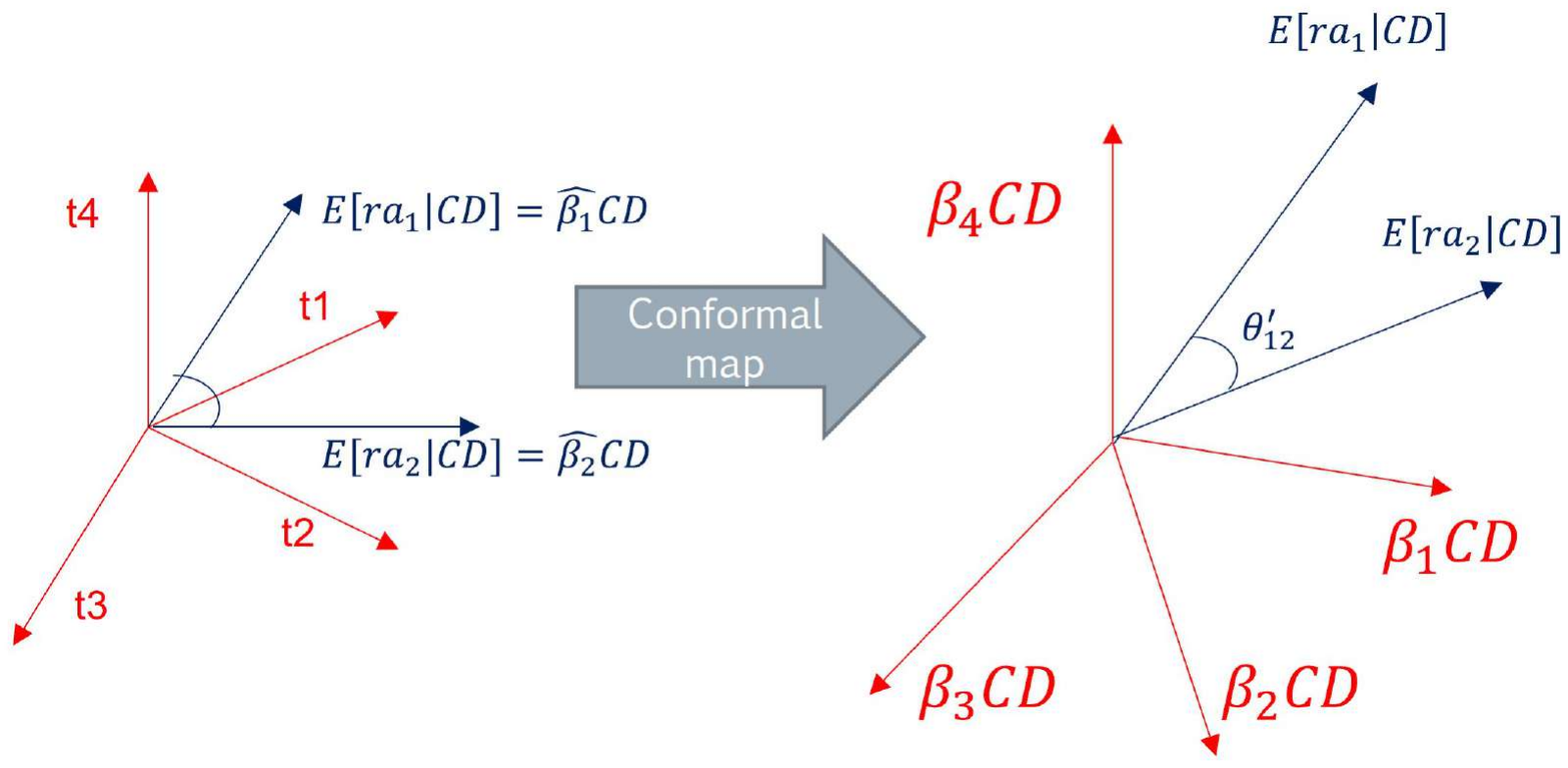}
	\caption{Conformal map between time-dimensional space and embedded space of sensitivities: $\theta_{12}^\prime$ is the angle in the new space with systematic and idiosyncratic components}
\end{figure}

In \citep{10.2307/3533246}, authors develop a representation theory for dynamic factor models. Our focus is on common drivers, which are different from common factors in that the former are sources of causality of dynamics (casual and persistent correlated relationships) and are not hidden data nor statistical information but publicly listed, whereas the latter is based on hidden statistical properties of the data. Moreover, our approach is based on projecting constituents into the space of sensitivities with respect to common drivers, as in other approaches in which constituents are projected into a space of common risk factors \citep{10.1016/j.eswa.2021.116308, 4744738}. Contrastingly, the space of sensitivities with respect to common drivers allows the inclusion of directionality in the search for diversification by incorporating information about returns dynamics and predictability conditional on publicly listed common drivers. Whereas common factors are based on the statistical properties and moments of past data.\par
Optimal common drivers selected by the commonality principle can refer to common factors in \citep{10.2307/3533246}, and share the same canonical decomposition:
\begin{equation}
    x_{it}=\ proj(x_{it}|\mathcal{G}(x))+\delta_{it}\ 
\end{equation}
Implying $\chi_{it}\ \in\ \mathcal{G}(x)$ and $\xi _{it}\ \bot\ \mathcal{G}(x)$, so that $\chi_{it}\ =\ proj(x_{it}|\mathcal{G}(x))$ and $\xi_{it}\ =\ \delta_{it}$. With Common Drivers(CD), $CD=\mathcal{G}(x)$, $\xi _{it}\ \bot\ CD$, $\chi_{it}\ =\ proj(x_{it}|CD) = E[x_{it}|CD]$. 
We show now with geometry how the canonical decomposition with common drivers preserves most idiosyncratic risk representation in the embedded spaces of sensitivities while adding systematic risk representation. In the mean-variance framework from MPT, the portfolio’s expected returns lie on a hyperplane of the constituents' expected returns, and portfolio risk lies on a hypersurface, as seen in Figure 2. The hyperplane is given by: 
\begin{equation}
E\left[r_p\right]=\sum_{i=1}^{n}w_iE[r_{a_i}]    
\end{equation} 

where, $E\left[r_p\right]$ are the portfolio’s expected returns, Tp is the Tangency Portfolio, $E\left[r_{a_i}\right]=\mu_i$ are the constituents' expected returns, as in Figure 2. $E\left[r_p\right]$ is linear in $E[r_{a_i}]$, portfolio risk $\sigma_p$ is non-linear in constituents' risk ${\sigma_{a_i}}$, and w are the weights, solution to the quadratic optimization in: 
\begin{equation}
\label{equation27}
w = \min_w{w^{T}\mathrm{\Sigma}\ w}
\end{equation}
with the tangency portfolio as the optimal solution. In Figure 3, we show the representation of portfolio constituents' expected returns for a period, in the time-dimensional space. Axis are points in time, $\theta$ are angles between expected returns, and the cosines are the correlations, ie, $\rho_{23}=\cos{\theta_2}$ is the correlation between $r_{a_2}$ and $r_{a_3}$ and $\theta_2=\hat{E[r_{a_2}]E[r_{a_3}]}$.
The expected portfolio returns conditional on the common drivers is a linear combination of constituents’ expected returns conditional on the same drivers: 
\begin{equation}
E\left[r_p|CD\right]=\sum_{i=1}^{n}w_iE[r_{a_i}|CD]    
\end{equation}
is a hypersurface. In Figure 4, we show the hypersurface and the tangency portfolio when solving the portfolio optimization. In Figure 5, on the left, we can see portfolio constituents' expected returns conditional on common drivers in a time-dimensional space with points in time as an axis. We demonstrate the existence of a conformal map from this space to another in which the conditional expectations are embedded in the space of sensitivities of constituents with respect to the common drivers (see the right side of Figure 5). In the embedded space, angles between conditional expected returns are a sum of two components, a systematic component, and an idiosyncratic component from the unconditional expectation case from MPT: 
\begin{equation}
\theta_{XY}^\prime={\alpha_1\theta}_{XY}+\alpha_2\gamma_{XY}\end{equation}
where $\cos{\theta_{XY}}$\ is $\rho_{XY}$ from $\mathrm{\Sigma}$\ \ in (\ref{equation27}), with ${\ \theta}_{XY}$ the idiosyncratic and $\gamma_{XY}$  the systematic component. The conformal map is such that the idiosyncratic component maintains the same proportion between angles in both spaces, and the idiosyncratic risk representation is kept at most in the new space. We will show next that the embedding of the time-dimensional space into a space of sensitivities with respect to common drivers is a conformal map.
\begin{lemma}
A Conformal map implies: 
\begin{equation}
\begin{split}
\theta_{XY}=f\left(t,\ E\left[r_{a_X}\right],E\left[r_{a_Y}\right]\right)=f\left(t,\ E\left[r_{a_X}\right](t),E\left[r_{a_Y}\right](t)\right)=\\
f(t)\Longrightarrow f\left(t,\ E\left[r_{a_X}|CD\right],E\left[r_{a_Y}\middle| C D\right]\right)=f\left(t\right)  
\end{split}
\end{equation}
\end{lemma}
If the angle in the mean-variance framework between unconditional expected returns is a function of both expected returns, which are also a function of time, the angle can be reduced to a function of time alone. A conformal map implies that the angle in the embedded sensitivity space, between the expected returns conditional on common drivers, which is a function of these conditional expectations and time, can be reduced to a function of time alone too. The map then is conformal at a given point in time.

\begin{proof}For the proof we analyze all possible cases.
\begin{itemize}
    \item Case I: Non-Common Drivers \begin{equation}
        \theta_{XY}=f(t,\ E\left[r_{a_X}|D_X\right],E[r_{a_Y}|D_Y])
    \end{equation}
    
\begin{equation}
\begin{split}
 \forall\ D_X,D_Y,D_X\neq D_Y:E\left[r_{a_X}|D_X\right]=\widehat{\beta_X}D_X, \widehat{\beta_X}=[\beta_{X_1},\ldots,\beta_{X_M}],\\ E\left[r_{a_Y}|D_Y\right]=\widehat{\beta_Y}D_Y, \widehat{\beta_Y}=\left[\beta_{Y_1},\ldots,\beta_{Y_M}\right]\\
 \Longrightarrow E\left[r_{a_X}|D_X\right]\bot E\left[r_{a_Y}|D_Y\right]
\end{split}
\end{equation}

\begin{equation}
 \left[\begin{matrix}E\left[r_{a_X}|D_X\right]\\E\left[r_{a_Y}|D_Y\right]\\\end{matrix}\right]=\left[\begin{matrix}\beta_{X_1},\ldots,\beta_{X_M},000,\ldots,0\\000,\ldots0,\beta_{Y_1},\ldots,\beta_{Y_M}\\\end{matrix}\right]\left[\begin{matrix}D_X\\D_Y\\\end{matrix}\right]
\end{equation}
Case I focuses on examples where drivers are not common. In this case, conditional expected returns are orthogonal in the embedded space of sensitivities with respect to drivers’ constituents, as seen in (34)-(36), which is not a valid case.
    \item 	Case II: Common Non-Casual Drivers \begin{equation}
        \theta_{XY}=f(t,\ E\left[r_{a_X}|D\right],E[r_{a_Y}|D])
    \end{equation} 
    \begin{equation}
    \begin{split}
      	E\left[r_{a_X}|D\right]=\widehat{\beta_X}D,\widehat{\beta_X}=\left[\beta_{X_1},\ldots,\beta_{X_M}\right],E\left[r_{a_Y}|D\right]=\widehat{\beta_Y}D, \\ \widehat{\beta_Y}=\left[\beta_{Y_1},\ldots,\beta_{Y_M}\right];\left[\begin{matrix}E\left[r_{a_X}|D\right]\\E\left[r_{a_Y}|D\right]\\\end{matrix}\right]=\left[\begin{matrix}\beta_{X_1},\ldots,\beta_{X_M}\\\beta_{Y_1},\ldots,\beta_{Y_M}\\\end{matrix}\right]\left[\begin{matrix}D_1\\\ldots\\D_M\\\end{matrix}\right]        
    \end{split}
    \end{equation}
\begin{equation}
	\left[\begin{matrix}E\left[r_{a_X}|D\right](t)\\E\left[r_{a_Y}|D\right](t)\\\end{matrix}\right]=\left[\begin{matrix}\frac{\partial r_{a_X}}{\partial D}\left(t\right)\ast D\\\frac{\partial r_{a_Y}}{\partial D}\left(t\right)\ast D\\\end{matrix}\right]
	\end{equation}
	if D is not cause of $r_{a_X}$ and \begin{equation}
	r_{a_Y}\Rightarrow r_{a_X}\left(t\right)\neq f\left(D\left(t\right)\right),r_{a_Y}\left(t\right) \neq f(D\left(t\right))
	\end{equation}
	Hence 
	\begin{equation}f\left(t,E\left[r_{a_X}|D\right](t),E\left[r_{a_Y}|D\right](t)\right)\neq f(t) \end{equation}
Case II focuses on common but non-casual drivers, in this case, if drivers are not the source of causality of constituents’ returns, these are not a function of the drivers, which is also not valid.	
	\item 	Case III: Common Causal Persistent Drivers (Commonality Principle):
	If D causes $r_{a_X}$ and $r_{a_Y}$:
	\begin{equation}\Rightarrow r_{a_X}\left(t\right)=f\left(CD\left(t\right)\right),r_{a_Y}\left(t\right) =f(CD\left(t\right))\end{equation}
	And we have: 
	\begin{equation}\theta_{XY}^\prime=f\left(t,\ E\left[r_{a_X}|CD\right],E\left[r_{a_Y}\middle| C D\right]\right)=f\left(t\right)
	\end{equation}
	embedding of t.
	\begin{equation}
	\begin{split}
	f\left(D_X\left(t\right)\right)=NN(D_X(t)\cong E[r_{a_X}|D_X(t)]\neq r_{a_X}(t);\\ f\left(D_Y\left(t\right)\right)=NN(D_Y(t)\cong E[r_{a_Y}|D_Y(t)]\neq r_{a_Y}\left(t\right);\\
	D_X\left(t\right)=D_Y(t) \iff D_X\left(t\right)=D_Y\left(t\right)=CD(t) 	    
	\end{split}
	\end{equation}
	Case III focuses on the commonality principle examples in which drivers are commonly casual and persistent. In this case, if common drivers (CD) cause constituents’ returns, these are functions of CD. If and only if those specific drivers are equal (commonality principle selection), the relationship between conditional expected returns is a function of time alone and the angle in the embedded space of sensitivities is also a function of time alone. Hence proving the conformal map.
\end{itemize}
\end{proof}

\section{Methodology}

The proposed portfolio optimization method can be summarized as follows. This methodology is carried out on every portfolio re-balance date (ie. when portfolio weights are updated). See Figure 6 for illustration:
\begin{itemize}
    \item Common drivers are selected as the top K (hyperparameter 1) with the highest common correlation with respect to all portfolio constituents using correlation time series daily returns and time windows $W_{CD}$ (hyperparameter 2).  
    \item With this common driver selection as inputs and one portfolio constituent as output, we train several feed-forward networks for the prediction task. We use time window $W_{NN}$ (hyperparameter 3) for many architectures (number of layers, neurons) and select the most accurate. We repeat this process for the rest of the portfolio constituents and the same common drivers.
    \item Once dynamics are approximated, for each previous optimal architecture for one portfolio constituent, we compute the sensitivities of each constituent with respect to the common drivers with Automatic Adjoint Differentiation (AAD) on feed-forward networks using the training set.
\item Each sensitivity is a function of the training set and we average it. Each constituent is assigned a vector of averages of sensitivity values with respect to the same common drivers and is used for embedding. A distance matrix is computed for all constituents with these averages of sensitivity values as coordinates.
\item This matrix is used for portfolio optimization first by finding the nearest positive semi-definite neighbor matrix with numerical methods, applying hierarchical clustering to the neighbor matrix in which hierarchies based on sensitivities are recorded. Finally, the positive semi-definite neighbor of the sensitivity matrix is sorted based on these hierarchies, and weights are computed based on the hierarchical partitions and clusters covariance matrices.
\end{itemize}

\subsection{Selection of Common Drivers}
\label{Subsection41}
From a set of M drivers with  $M >> N$, N being the number of constituents of the portfolio. For each constituent, we rank correlations with respect to all drivers for different lags and time horizons, with a threshold that depends on the lag. We select drivers that have passed the thresholds the greatest number of times among all portfolio constituents. We now show the algorithm: \par

\begin{equation}\forall\ {Asset}_i,{SD\ }_i\ \ i=1,\dots N,\ {\forall\ Driver}_j\ j=1,\dots M
\label{equation31}
\end{equation}

\begin{equation}
\begin{split}
 \ 
{(Driver}_j\in\ {SD\ }_i)\Rightarrow\\ (corr\left(\ {Asset}_i\left(t\right),\ {Driver}_j\left(t\right)\right)>T0\\
\wedge \ corr\left(\ {Asset}_i\left(t\right),\ {Driver}_j\left(t-1\right)\right)>T1)   
\label{equation32}
\end{split}
\end{equation}

\begin{equation}
A=A_0,\ B=\ \left\{b\in A_i:b\geq a\ \forall a\ \in A_0\ \right\},\ A_1=\ A_0\setminus B_1, 
\label{equation33}
\end{equation}

\begin{equation}
B_{i+1}\ =\ \left\{b\in A_i:b\geq a\ \forall a\ \in A_i\ \right\},\ A_{i+1}=\ A_i\setminus B_{i+1}
\label{equation34}
\end{equation}

\begin{equation}
\begin{split}
 (A={{Driver}_j}| \max( \#({Driver}_j\ \in \ {SD\ }_i)\\ 
 \forall\ {Driver}_j\ , {SD\ }_i \ i=1,\dots N,\ \ j=1,\dots M)   
 \label{equation35}
\end{split}
\end{equation}
 
\begin{equation} B_k={CD}_i,\ k=\max{\left(\#CD\right)},\ i=1,\dots N
\label{equation36}
\end{equation}

Equation (\ref{equation32}), states that, for a driver to be a specific driver for a particular constituent, it must have correlations above thresholds T1 and T0 for respective lags 1 and 0 (hyperparameters 4 and 5). Equations (\ref{equation33}) and (\ref{equation34}) are the formulations for the problem of finding the set of (i+1) elements that have a greater value than a threshold from other sets of elements. Equation (\ref{equation35}) is adapting Set A to our problem because we want, from all drivers of the drivers set, those that are simultaneously specific for the greatest number of portfolio constituents. K is the hyperparameter 1 of choice that indicates the maximum number of common drivers to select for the model implementation. $B_k$ will be the k common drivers optimally chosen. Optimal in terms of passing the thresholds and being repeated the maximum number of times among portfolio constituents. Figure 6, the blue bordered box shows this step of the method.

\subsection{Optimal network selection}
\label{subsection421}
For each portfolio constituent, we fit a neural network with the constituent as output, and the same common drivers as inputs, using time series daily returns data from the past period we want to approximate the dynamics and extract sensitivities. We fit many architectures based on different configurations in terms of the number of layers, and neurons. Fitting window lengths, hyperparameter 3, $W_{NN}$, and lags between outputs and inputs (hyperparameter 6). 
We train the architectures on the fitting period and evaluate the fit with an error measure like Mean Square Error (MSE). We select the optimal architecture based on this metric. We used a multi-layer perceptron. Finally, each portfolio constituent will have an optimal architecture. In Figure 6, the red-bordered box shows this step of the method.

\begin{figure}
\vspace{-25mm}
	\centering
	\includegraphics[width=180 mm]{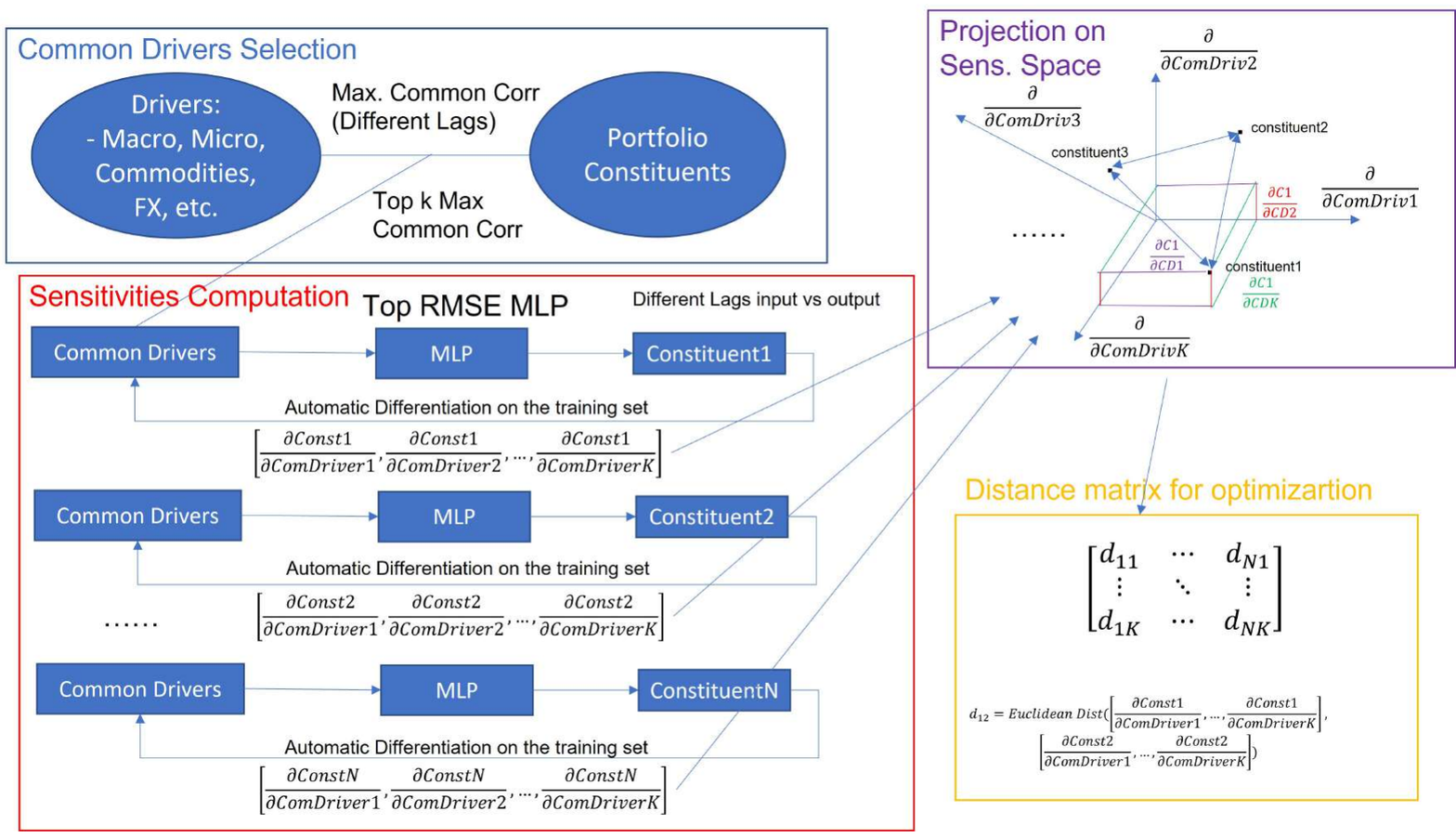}
	\caption{Methodology modules: sensitivities are extracted from multi-layer perceptrons (MLP) with AAD and used as coordinates in the projective space, each constituent has its own vector. Distance matrix computed in this space.}
\end{figure}

\subsection{Sensitivity Analysis}
For each portfolio constituent, an optimal architecture is selected, and AAD with TensorFlow and GradientTape is used to approximate the partial derivatives of the constituent with respect to the common drivers (sensitivities), as in \citep{huge2020differential} where authors apply this to derivative pricing PDEs in the vanilla net case. These sensitives are discrete functions with a value for each time step of the fitting period. To obtain a metric of each sensitivity for that period, we use the average value. However, the choice of different functions to resume this sensitivity information from the fitting period can improve the method's performance. 	
For each portfolio constituent $a_i$:
\begin{equation}
\begin{split}
 {E\left[r_{a_i}\middle| C D\right]=y}_i\left(t\right)= F\left(\frac{\partial y_i}{\partial x_1}(t),\frac{\partial y_i}{\partial x_2}(t),\ldots,\frac{\partial y_i}{\partial x_M}(t),\frac{\partial x_1}{\partial t},\ldots,\frac{\partial x_M}{\partial t},\frac{\partial y_i}{\partial t},x_1,\ldots,x_M\right)   
\end{split}
\end{equation}
is approximated by a neural network $({NN}_i)$, and $\frac{\partial y_i}{\partial x_1}(t),\frac{\partial y_i}{\partial x_2}(t),\ldots,\frac{\partial y_i}{\partial x_M}(t)$, sensitivities with respect to the common drivers are obtained via AAD. Sensitivities are the partial derivatives defined as: ${s_{ij}|}_{x_n}^{{NN}_i}=\frac{\partial y_i}{\partial x_j}(x_n)$, which refers to the sensitivity of the output of the neuron in the output layer of ${NN}_i$\ respect to the input of the $j^{th}$\ neuron in the input layer evaluated in the sample $x_n$. It is obtained by applying the chain rule to the partial derivatives of the inner layers. 

\subsection{Sensitivity Distance Matrix}
Once all sensitivities are obtained for all portfolio constituents with respect to the common drivers, we need to incorporate them into the problem of portfolio optimization. We represent all portfolio constituents into a K-dimensional space, with K (hyperparameter 1) being the number of common drivers (same for all), and coordinates being the averages of sensitivity values of the portfolio constituents with respect to each of the common drivers, as in Figure 7.\par

\begin{figure}
\label{Image7}
	\centering
	\includegraphics[width=95mm]{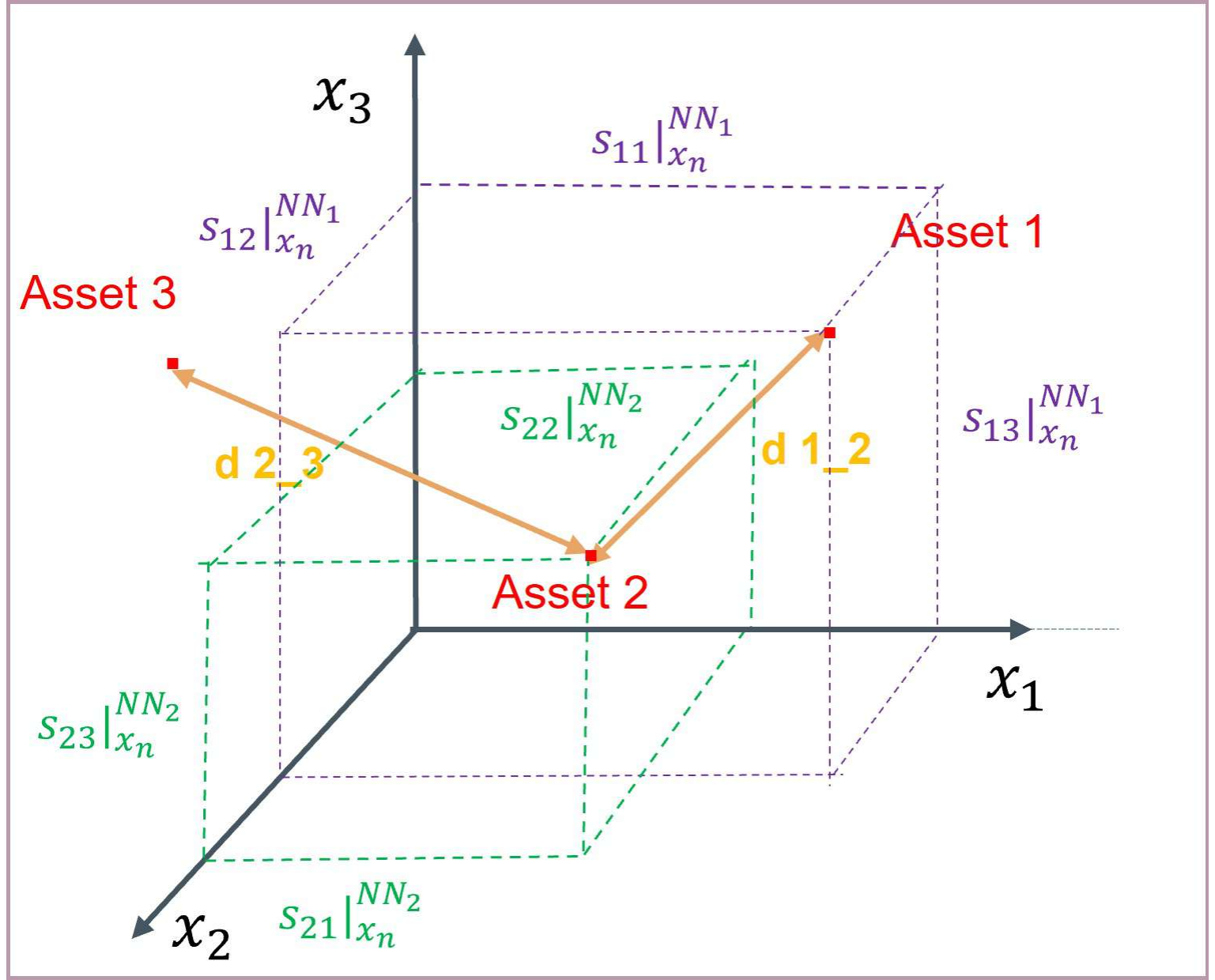}
	\caption{Embedded space of sensitivities: sensitivities as coordinates and the distance metric in orange}
\end{figure}

Portfolio constituents, $y_1,\ldots,y_N$ are embedded in the space of averages of sensitivity values with respect to common drivers:
\begin{equation}
 	\left[\begin{matrix}y_1\\\vdots\\y_N\\\end{matrix}\right]=\left[\begin{matrix}\frac{\partial y_1}{\partial x_1}(t)&\cdots&\frac{\partial y_1}{\partial x_M}(t)\\\vdots&\ddots&\vdots\\\frac{\partial y_N}{\partial x_1}(t)&\cdots&\frac{\partial y_N}{\partial x_M}(t)\\\end{matrix}\right]\left[\begin{matrix}x_1\\\vdots\\x_M\\\end{matrix}\right]\cong\left[\begin{matrix}{s_{11}|}_{x_n}^{{NN}_1}&\cdots&{s_{1M}|}_{x_n}^{{NN}_1}\\\vdots&\ddots&\vdots\\{s_{N1}|}_{x_n}^{{NN}_N}&\cdots&{s_{NM}|}_{x_n}^{{NN}_N}\\\end{matrix}\right]\left[\begin{matrix}x_1\\\vdots\\x_M\\\end{matrix}\right]   
\end{equation}

In this space of common drivers’ averages of sensitivity values coordinates in which all portfolio constituents are represented we compute a distance matrix.

\begin{definition}	Sensitivity Matrix:
	Distance matrix between portfolio constituents in the embedded space of functions of the sensitivity values.
	\begin{equation}
	\begin{split}
	S=\left[\begin{matrix}s_{11}&\cdots&s_{1N}\\\vdots&\ddots&\vdots\\s_{N1}&\cdots&s_{NN}\\\end{matrix}\right],\\ s_{ij}=d\left(\frac{\partial y_i}{\partial x},\frac{\partial y_j}{\partial x}\right)=d\left(\vec{\beta_i},\vec{\beta_j}\right)=d\left(\left[\beta_i^1,\ldots,\beta_i^M\right],\left[\beta_j^1,\ldots,\beta_j^M\right]\right)	    
	\end{split}
	\label{sensmatrix}
\end{equation}
Coordinates of portfolio constituents are the functions (average) of sensitivity values with respect to the common drivers in the training dataset, and the distance matrix is called the Sensitivity matrix.
\end{definition}

Sensitivities include optimal causal and persistence information of constituents’ dynamics for the fitting period. By embedding constituents into a sensitivity space, the similarity matrix (sensitivity matrix) allows us to compare portfolio assets based on their casual (and persistent) common dynamics. This common causal structure in which constituents are projected allows the geometric link, the conformal map, to preserve idiosyncratic risk representations at most from a traditional MPT setup, with no exogenous information, while adding systematic diversification with exogenous drivers. Additionally, our setup allows for directional diversification with those projections, in which risk diversification has an additional dimension with the true dynamics information of portfolio returns with respect to public and listed variables, not statistical factors, consequently adding an improved approximation of directionality in portfolio returns for risk diversification. \par

\subsection{Hierarchical Sensitivity Parity: Convex Optimization and Hierarchical Clustering}

To solve the convex optimization problem for diversification by means of the sensitivity matrix, we cannot use quadratic optimizers. By applying hierarchical clustering constituents are clustered based on hierarchies from portfolio dynamics information with the aforementioned benefits. As \citep{Prado2016BuildingDP} applies a hierarchical clustering to the correlation matrix to improve diversification with the hierarchical representation of projections from assets between one and another, we can extract useful information from the hierarchical representation of projections into the space of sensitivities with respect to common drivers, which already has directionality benefits, and improve diversification upon it. We find the performance is improved if we use a nearest positive-definite neighbor of the sensitivity matrix computed as in \citep{HIGHAM1988103}. Then we apply single-linkage clustering to the neighbor matrix:
\begin{equation}
\begin{split}
D(X,Y)=\min_{x\in X, y\in Y}d(x,y)\\
d(x,y)=\sqrt{\sum_{n=1}^{N}\left(S_{n,x}-S_{n,y}\right)^2} 
\end{split}
\label{singlelink}
\end{equation}
with S like (\ref{sensmatrix}), or the neighbor \citep{HIGHAM1988103} of S, so that the single-linkage algorithm is applied directly to S (\ref{sensmatrix}), or a positive semi-definite neighbor matrix of S. Then, clustered items from the single-linkage algorithm are sorted by distance (\ref{singlelink}). Portfolio constituents will end up close to those similar to them and far from very different ones using a distance metric (\ref{singlelink}) based on the sensitivity matrix (\ref{sensmatrix}) or a positive semi-definite neighbor matrix. Finally, we apply Recursive Bijection from \citep{Prado2016BuildingDP} to compute the weights. The sorted distance matrix is used to explore the clustering tree from top to bottom and on each partition two clusters compete for the weights. For each cluster the weights we will use to compute its volatility are \citep{Prado2016BuildingDP}:

\begin{equation}
    w=\frac{diag\left[V\right]^{-1}}{trace(diag\left[V\right]^{-1})}
\end{equation}
 
V is the covariance matrix of the constituents of the cluster. This way we can compute the variance of both clusters as:
\begin{equation}
\sigma_1={w_1}^TV_1w_1
\sigma_2={w_2}^TV_2w_2
\end{equation}
Now we compute two factors that we will use to re-escalate the weights of the clusters, $\alpha_1$ and $\alpha_2$.
\begin{equation}
\alpha_1=1-\frac{\sigma_1}{\sigma_1+\sigma_2};\alpha_2=1-\alpha_1
\end{equation}
so $w_1=w_1\alpha_1$, $w_2=w_2\alpha_2$. "Suppose we have $2^n$ assets, then our initial vector w will be a vector of $2^n$ ones, with a sum of $2^n$. In the first step, we multiply half the vector by $\alpha_1$ and the other half by $\alpha_2=1-\alpha_1$. If we make pairs with an updated weight from the first half and another one from the second, each pair has a sum of 1 and there are $\frac{n}{2}$ pairs, meaning that the total sum has been halved in this step. Recursively, in every step we take a vector of identical weights and multiply half of it by $\alpha_i$ and the other half by $1-\alpha_i$ , halving their sum. If we consider that we visit all of our assets in every step, the whole process will take $log_{2}(2^n) = n$ steps and the final sum will be $2^n(\frac{1}{2})^n = 1$", as mention in \citep{HRPmiker}.\par

We apply recursive bisection, but with a clustering tree obtained from the single-linkage algorithm using a distance metric (\ref{singlelink}) of the sensitivity matrix (\ref{sensmatrix}) or a positive semi-definite neighbor matrix. In contrast to Hierarchical Risk Parity (HRP) from \citep{Prado2016BuildingDP}, where the author applies recursive bisection to a clustering tree obtained from the single-linkage algorithm using a distance metric of the correlation matrix. Hence our method is called Hierarchical Sensitivity Parity (HSP).\par

\subsection{Algorithm and model configuration for the experiments}
\IncMargin{1em}
\begin{algorithm}[H]
\SetAlgoLined
\KwData{$Constituents$: Daily time series returns of portfolio constituents\;
$Drivers$: Daily time series returns of the universe of driver candidates\;
$Dates$: Array of portfolio re-balance dates\;
$CommonDriversDate$: Array of common drivers selection dates\;
}

\KwResult{$W$ dictionary with re-balancing dates as Key, and weights for each portfolio constituents as Values}
CD=CommonDriversSelection(Constituents, Drivers)\;
    \ForEach{$Date$ in Dates}{

        \If{$Date$ in CommonDriversDate}{
        CD=CommonDriversSelection(Constituents, Drivers)\;
        }
        s = []\;
        \ForEach{Constituent in $Constituents$}{
            NN = OptimalArchitecture(Constituent,CD)\;
        Sens = AAD(NN, Constituent, CD)\;
        Sens = SensFunction(Sens, "Average")\;
        s = append(Sens)\;
        }
        SensitivityMatrix = distance(s, s; "Euclidean")\;
        B = nearestPD(SensitivityMatrix)\;
        link = linkage(B)\;
        sortedIndex = Sorting(link)\;
        PCreturns = Constituents[(Date-3months):Date]\;
        CovarianceMatrix = Covariance(PCreturns)\;
        $W[Date]$ = RecursiveBisection(CovarianceMatrix, sortedIndex)\;
    }
\caption{Hierarchical Sensitivity Parity (HSP)}\label{HSP}
\end{algorithm}

Algorithm \ref{HSP}, shows the pseudo-code for Hierarchical Sensitivity Parity(HSP). CommonDriversSelection() implements the algorithm from subsection \ref{Subsection41} and returns the common drivers set on each date of drivers’ selection. OptimalArchitecture() returns the best neural network model as in subsection \ref{subsection421}. AAD() returns the sensitivities of each portfolio constituent with respect to the current common drivers for the fitting window. SensFunction() returns the average of sensitivity values. A table with all averages of sensitivity values, s, is used to compute the pair-wise euclidean distance (sensitivity matrix) with distance() function. nearestPD(), returns the nearest positive semi-definite matrix from an input matrix as in \citep{HIGHAM1988103}. linkage(), which returns the hierarchical clustering encoded as a linkage table describing the order in which clusters were formed. Sorting(), returns the order implied by the linkage matrix (sortedIndex) to sort the distance matrix. RecursiveBisection(), is the recursive bisection function from the HRP method in \citep{Prado2016BuildingDP} which returns the portfolio weights given a covariance matrix and an array of matrix indexes. In algorithm \ref{HSP}, the sortedIndex comes from
the single-linkage algorithm using a distance metric (\ref{singlelink}) of the sensitivity matrix (\ref{sensmatrix}) or a positive semi-definite neighbor matrix, instead of single-linkage algorithm using a distance metric of the correlation matrix of portfolio constituents' historical returns as in HRP  \citep{Prado2016BuildingDP}. For the experiments, CommonDriversDate is a dates array on a six-month basis, with drivers left fixed for subsequent weights re-balancing dates. Dates is a monthly basis dates array for the re-balancing (weights update) dates.\par

\section{Implementation}
We use Bloomberg as our data source for two sets of data stored in a database:
\begin{itemize}
    \item The drivers’ dataset consists of 1200 time series data from 2012 to the end of 2021 including spot and option prices for different strikes and tenors for the main Foreign Exchange (FX) crosses in the world (FX information); government bonds with different maturities from the most important countries in all world economies including the five continents (rates information); main macroeconomic indicators (monthly and quarterly series) for the most important countries in terms of economy in the five continents (information about the economies); main equity indexes across all geographies; mutual funds indexes for government, corporate, investment grade, and high yield debt for Europe, USA, Asia and emerging markets (fixed income markets); Credit Default Swap (CDS) price data and credit mutual fund indexes for investment grade and high yield for Europe and USA (credit markets); futures’ prices for the main commodities traded in the world; smart beta Exchange Traded Funds (ETFs) representing the main risk factors in the market and literature; main crypto assets; equity indexes option implied volatility time series for different strikes and tenors (market risk indicators) and the equity sectors’ ETFs for the USA and Europe. This sample is a summary of all public information and traded products available for practitioners in the financial industry, representing the financial markets. These can serve as driver candidates. 
    \item The second dataset consists of two sets of 14 stocks’ daily time series price data from Stoxx 600 and SP500 from different sectors, in order to start with some level of diversification in the sample.  We can see the two sets in Table 1.
    \item Time series are converted to percentage change (returns) in both datasets when performing the common drivers’ selection algorithm, as well as the rest of the methodology.
\end{itemize}
    
\begin{table}[!hbt]
\centering
\begin{tabularx}{\columnwidth} {@{} l*{3}{X} @{}}
\toprule
SXXP (Portfolio EU) & SPX (Portfolio USA)\\
\midrule
ASML HOLDING NV      & GENERAL ELECTRIC CO         \\
LVMH MOET HENNESSY LOUIS        & GOLDMAN SACHS GROUP INC     \\
SAP SE       & APPLE INC                   \\
SIEMENS AG-REG      & NVIDIA CORP                 \\
L'OREAL        & DOVER CORP                  \\
SANOFI       & FORD MOTOR CO               \\
ALLIANZ SE-REG       & ORACLE CORP                 \\
SCHNEIDER ELECTRIC SE        & PACKAGING CORP OF AMERICA   \\
TELEFONICA SA       & MCDONALD'S CORP             \\
BANCO SANTANDER SA       & PFIZER INC                  \\
INTL CONSOLIDATED AIRLINE-DI      & SCHLUMBERGER LTD            \\
REPSOL SA      & BLACKROCK INC               \\
INDRA SISTEMAS SA     & PHILIP MORRIS INTERNATIONAL \\
	GRIFOLS SA      & EQUINIX INC \\
\bottomrule
\end{tabularx}
\caption{Portfolios for experiments: One for Europe, other for USA markets}
\end{table}

We perform experiments for different values of the hyperparameters. We use K (hyperparameter 1) as the number of common drivers with values 10, 15, 20, and 30; we use both 6 and 12 months of daily time series data for the Correlation window $W_{CD}$ (hyperparameter 2) in optimal driver selection; neural networks training window $W_{NN}$ (hyperparameter 3) 60, 90, and 125 market days and for the correlations' thresholds T1 and T0 for respective lags 1 and 0 (hyperparameters 4 and 5) for common drivers’ selection we use a ranking rule in which we pick the top K (hyperparameter 1) candidates most correlated to the greatest number of portfolio constituents. To do this, we sum up all correlations for each driver candidate with respect to all constituents and pick the top K sums. Another approach could be to calibrate T1 and T0 to model performance or to select T1 and T0 so that the number of candidates passing these thresholds matches K, or the number of candidates matches K and are appropriate. This is because we may have one stock in the portfolio that is already part of an equity or sector index which is a driver candidate and we do not want to add that driver, so the user can drop it and have more candidates to choose from due to the threshold. We use lags of 0,1,2,5,10, and 20 between outputs and inputs (hyperparameter 6) in the neural networks. \par
We have two possible versions for our method HSP: SELECT means we make adjustments to the final drivers’ selection if there is some information from a stock already in the driver (index), if we think the correlation is spurious (not related at all), or if there is high multicollinearity in the set of drivers; OPT means we leave the algorithm to select the top K drivers in terms of highest correlation values with no adjustments. For the driver selection, we can use lag 0, or 1, separately or both simultaneously, giving different sets of optimal drivers and performances (more than 1 lag is usually useless). Portfolio optimization (re-balance) is carried out once a month and weights are left fixed for the subsequent month. We perform experiments with all modules including driver selection carried out on each re-balance date, and also experiments in which we perform common drivers’ selection on a 6-month basis, leaving this selection fixed for the following 6 portfolio re-balances, implementing the rest of the modules with the same set of drivers ( neural network selection, training, sensitivities extraction, and HSP algorithm) until the next common drivers’ selection date. \par
All back-test performances are measured out-of-sample, all modules, models, and computations are made with past data, and solutions are left fixed for present and future portfolio decisions. The same applies to all competitors' methods for consistency. The focus is on portfolio optimization so equity names are left fixed for all methods, and we restrict all methods to have weights between 3
We have included all mean-variance optimization techniques (Maximum Sharpe, Minimum Volatility, Quadratic Utility, Target Return, etc), and the HRP method from \citep{Prado2016BuildingDP} to test the hypothesis that we can improve performance by adding hierarchical sensitivity dynamics information with all the aforementioned benefits. This is in contrast to using hierarchical correlation information as in \citep{Prado2016BuildingDP},  which is one of our contributions. Indirectly, we are comparing to other methods that use risk factors instead of drivers when we perform experiments forcing us to select smart beta ETFs or equity and sector indexes as the common drivers and we contrast performance with our method for optimal common driver selection. We test that adding risk factors as common drivers produce lower performance. All methods use the same past window for the covariance or correlation matrices for optimization including recursive bisection for HRP and HSP.\par
Due to space limitations and graph clarity, we only ever show the top performers, which always includes our method, from a multitude of experiments, but the ones that we do not show are underperforming these ones, many by far.\par

Next, we show experiments for the USA and Europe portfolios with performance for all methods measured with Net Asset Values (NAVs), which consist of the time series daily dollars under management if the strategy starts with 100 dollars, computed as if they all started on 01/06/2020 and ended on 01/12/2021. For our method, there are three dates where common drivers are selected: 01/06/2020, 01/01/2021, and 01/07/2021. All re-balances are performed on the 1st of each month for all methods and weights are kept fixed for the next 30 days. After this, we focus on long-term investments and perform the same experiments, but over 6 years and with all modules including portfolio selection carried out in each re-balancing date monthly. \par

\subsection{Portfolio USA}
\label{USAcase}
We perform experiments for all methods including 1/N (equal-weighted), with different model constraints (ie. for target return) and a maximum of 10\% weight for any constituent (1/N = 7,1\%) to avoid concentration. NAVs are computed with an initial investment of 100, from 01/06/2020 up to 01/12/2021. \par

\begin{figure}
	\centering
	\includegraphics[width=110 mm]{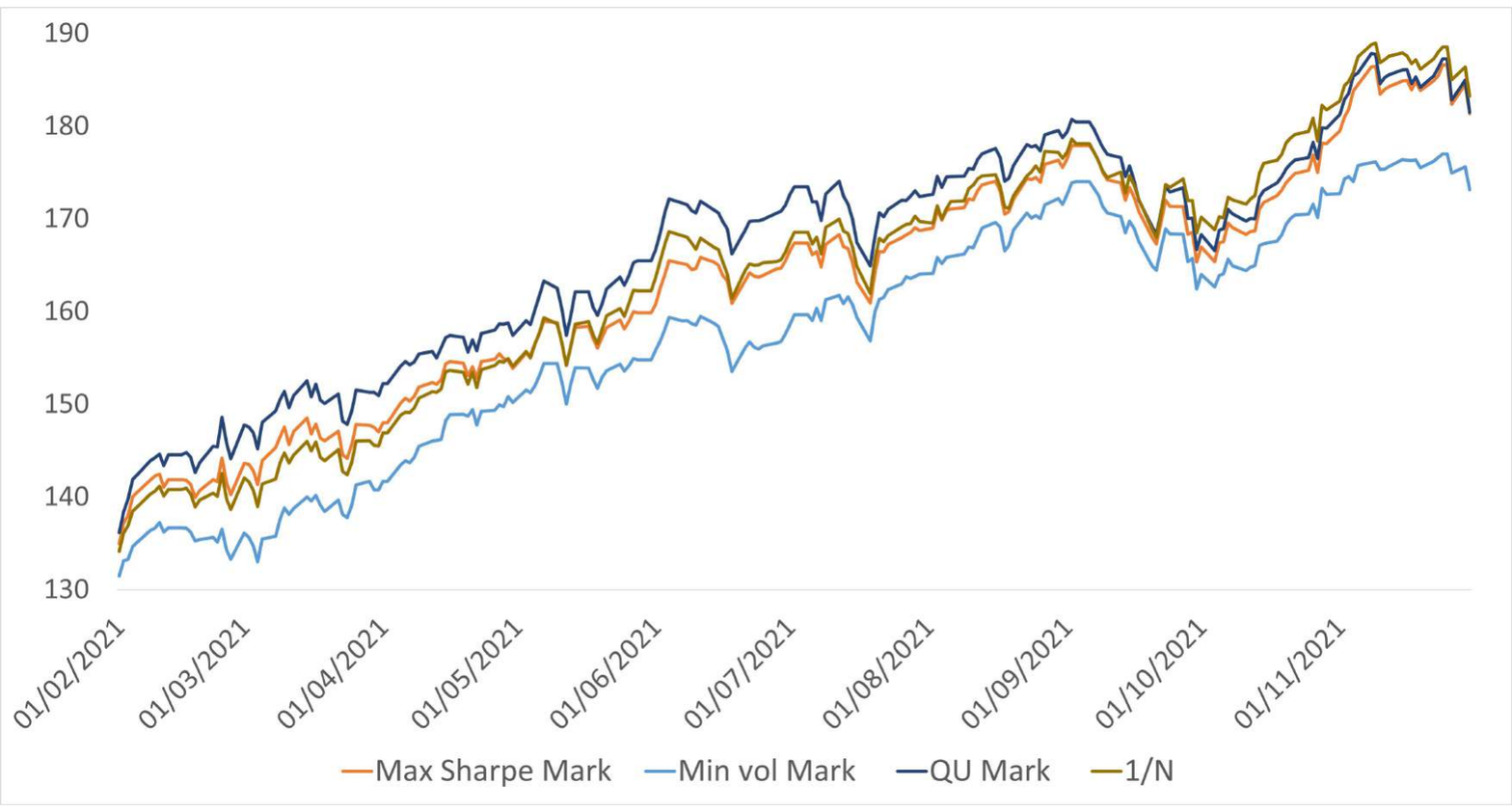}
	\caption{NAVs for USA portfolio top mean-variance methods and 1/N: NAV starting from 01/06/2020, showing subperiod (02/2021-12/2021).}
	\label{figure5}
\end{figure}

In Table \ref{table2}, we show returns, volatilities (annualized), and Sharpes for the mean-variance winner methods and 1/N only, for the entire period from 01/06/2020 to 01/12/2021. In Figure \ref{figure5}, we show NAVs from 02/2021 to 12/2021 just for them.\par 

\begin{table}[!hbt]
\centering
\begin{tabularx}{\columnwidth} {@{} l*{4}{X} @{}}
\toprule
    & Max Sharpe(Mark) & Min Vol(Mark) & QU Mark & 1/N   \\
\toprule
\midrule
Return    & 49\%            & 44\%         & 49\%    & 50\%  \\
Vol (Ann) & 16\%            & 15\%         & 17\%    & 17\%  \\
Sharpe    & 3,061           & 3,002        & 2,920   & 3,006 \\
\bottomrule
\end{tabularx}
\caption{USA portfolio performance metrics for top mean-variance methods and 1/N: Returns, Risks, and Sharpes for full period: 01/06/2020 – 01/12/2021}
\label{table2}
\end{table}

Now we include our method (HSP for results): On table \ref{table3}, we include the previous top mean-variance performers, Max Sharpe, with the 1/N, and HRP from \citep{Prado2016BuildingDP}. We also include the best-performing versions of our method, HSP 6m LAG1 SELECT, HSP 6m LAG0 OPT, HSP 6m LAG1 OPT, with the three out-performing the rest of the methods. We use 6 or 12 months past data windows for correlations for driver selection with the algorithm from Section \ref{Subsection41} (hyperparameter 2). Correlation thresholds T1 and T0 for respective lags 1 and 0 (hyperparameters 4 and 5) are selected so that on all drivers’ selection dates we have enough candidates and the number of constituents is similar to the number of common drivers (hyperparameter 1). In the next section, we further explain the selection of hyperparameters 1, 2, and 3. We select the optimal neural network architecture for fitting and sensitivity computation as in Section \ref{subsection421} from a set of neural networks with inputs and outputs having a lag 0, or lag 1, or both lags simultaneously (hyperparameter 6). We use different neural network fitting windows (hyperparameter 3), 60, 90, and 125 market days, for each of the previous sets to find the optimal architecture.\par 
HSP 6m LAG1 OPT means a 6-month window for correlations driver selection (hyperparameter 2), and only lag 1 between constituents and drivers for the set of neural networks to find the optimal as in Section \ref{subsection421} (hyperparameter 6). OPT means the full algorithmic selection based on thresholds for correlation values as in Section \ref{Subsection41} is used. The SELECT case means the common drivers’ selection from the algorithm in Section \ref{Subsection41} has been tuned based on spurious correlation, stock as part of an index driver, or multicollinearity. In this case, the ranking of the most commonly correlated (based on the commonality principle) is used taking these restrictions into account. The number of common drivers varies in each selection date as correlations vary a lot, so hyperparameter 1 depends on hyperparameters 4 and 5, the correlation thresholds, tuned accordingly. In Section management insights we explain how to improve performance and select between the HSP versions. \par

\begin{figure}
	\centering
	\includegraphics[width=130 mm]{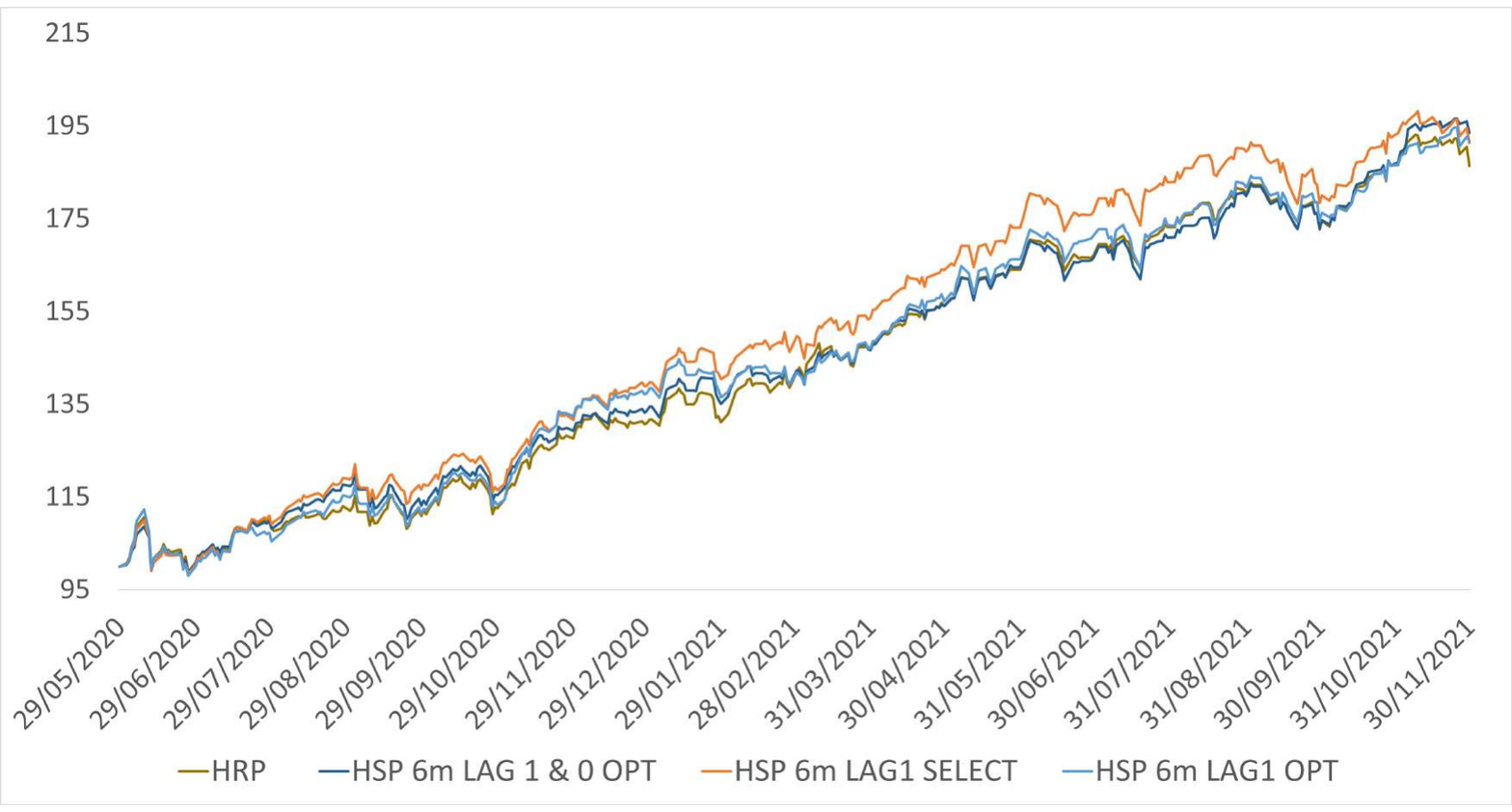}
	\caption{NAVs for USA portfolio for top mean-variance methods, 1/N, HRP, and HSP for different model hyperparameters: NAV starting from 01/06/2020. Showing the top 4 performers}
	\label{figure6}
\end{figure}

\begin{figure}
	\centering
	\includegraphics[width=130 mm]{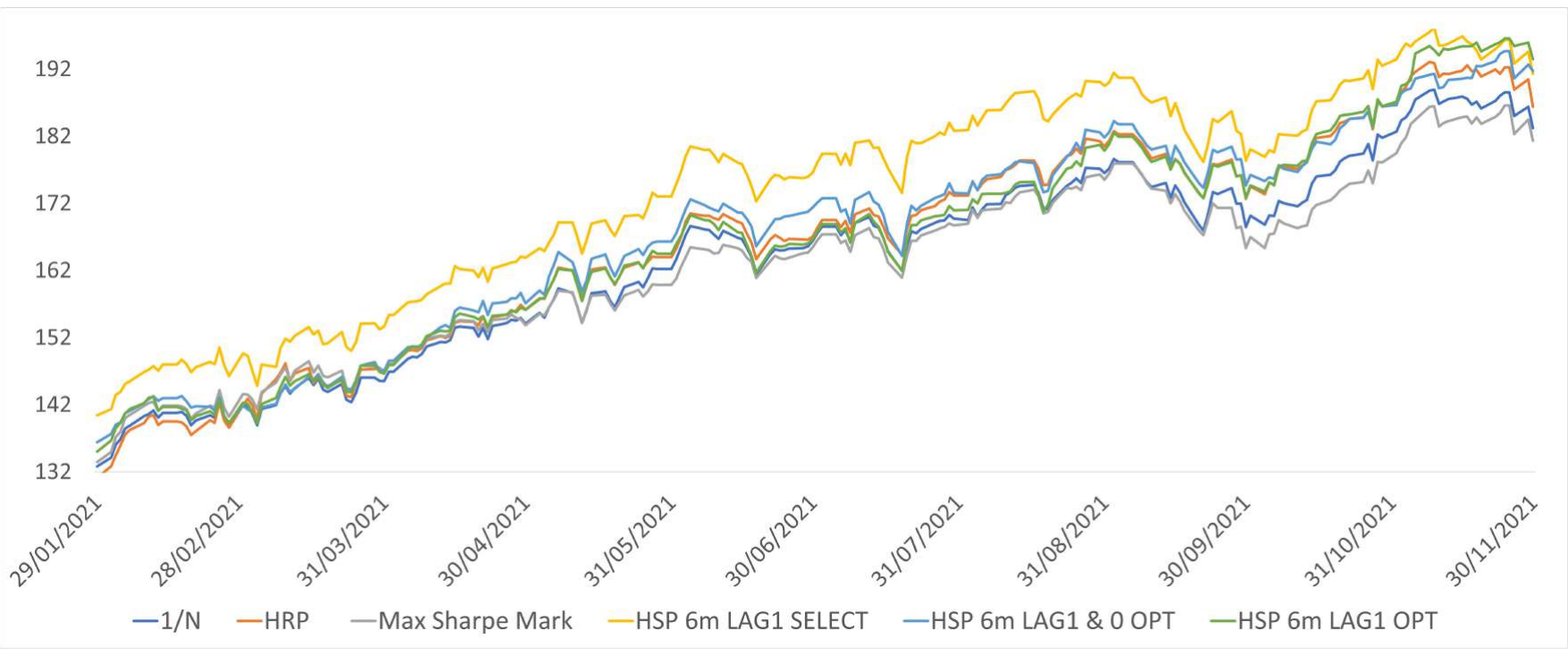}
	\caption{NAVs for USA portfolio for top mean-variance methods, 1/N, HRP, and HSP for different model hyperparameters: NAV starting from 01/06/2020, zoom of Figure \ref{figure6} showing subperiod from 01/2021}
	\label{figure7}
\end{figure}

In Figure \ref{figure6}, we show NAVs for the entire period, and in Figure \ref{figure7}, a sub-period from 01/02/2021 to 01/12/2021. It is relevant to mention that the last common driver selection is on 01/07/2021, 5 months before the last underperforming for all models that we can see in November 2021 onward (see Figure \ref{figure7}). It is quite likely that 5 months after the last common drivers update this optimal selection could have changed, and if we had updated them more often than every 6 months, it is probable that we would not see that loss in our model. As can be seen in Subsection 5.3 with better long-term performance by adding more frequent common drivers’ updates. In any case, all models’ performance fall, except HSP 6m LAG 0 OPT.\par

\begin{table}[!hbt]
\centering
\begin{tabularx}{\columnwidth} {@{} l*{5}{X} @{}}
\toprule
          & HSP 6m LAG 1 SELECT & HSP 6m LAG 0 OPT & HSP 6m LAG 1 OPT & 1/N  & HRP \\
\toprule
\midrule
Return    & 54\%                & 55\%             & 54\%             & 50\% & 52\%               \\
Vol (Ann) & 17\%                & 17\%             & 17\%             & 17\% & 17\%               \\
Sharpe    & 3,157               & 3,340            & 3,116            & 3,0  & 2,954              \\
\bottomrule
\end{tabularx}
\caption{USA portfolio performance metrics for 1/N, HRP, HSP for different model hyperparameters: Returns, Risk and Sharpes for full period: 01/06/2020 – 01/12/2021}
\label{table3}
\end{table}

\subsection{Portfolio EU}
\label{EUcase}

We keep all the same for the EU experiments as for the US case, the only thing it changes is the equity names. Our method is still the best performer in the EU case. In Table \ref{table4}, we show the top mean-variance case, 1/N, HRP, and two choices from our method, HSP 6m LAG 1 OPT, which means a 6-month window for hyperparameter 2, LAG 1 for the hyperparameter 6. OPT means full algorithmic selection based on thresholds for correlation values, hyperparameters 4 and 5, selected as in the USA case giving hyperparameters 1. 
HSP 6m LAG 0 \& 1 SELECT, where again, SELECT means some common drivers’ candidates that pass hyperparameters 4 and 5 thresholds are changed as already described, the set of neural networks for optimal architecture contains both lags 0 and 1 for hyperparameter 6. Performances on Table \ref{table4} are for the entire period, 01/06/2020 – 01/12/2021. \par

\begin{figure}
	\centering
	\includegraphics[width=120 mm]{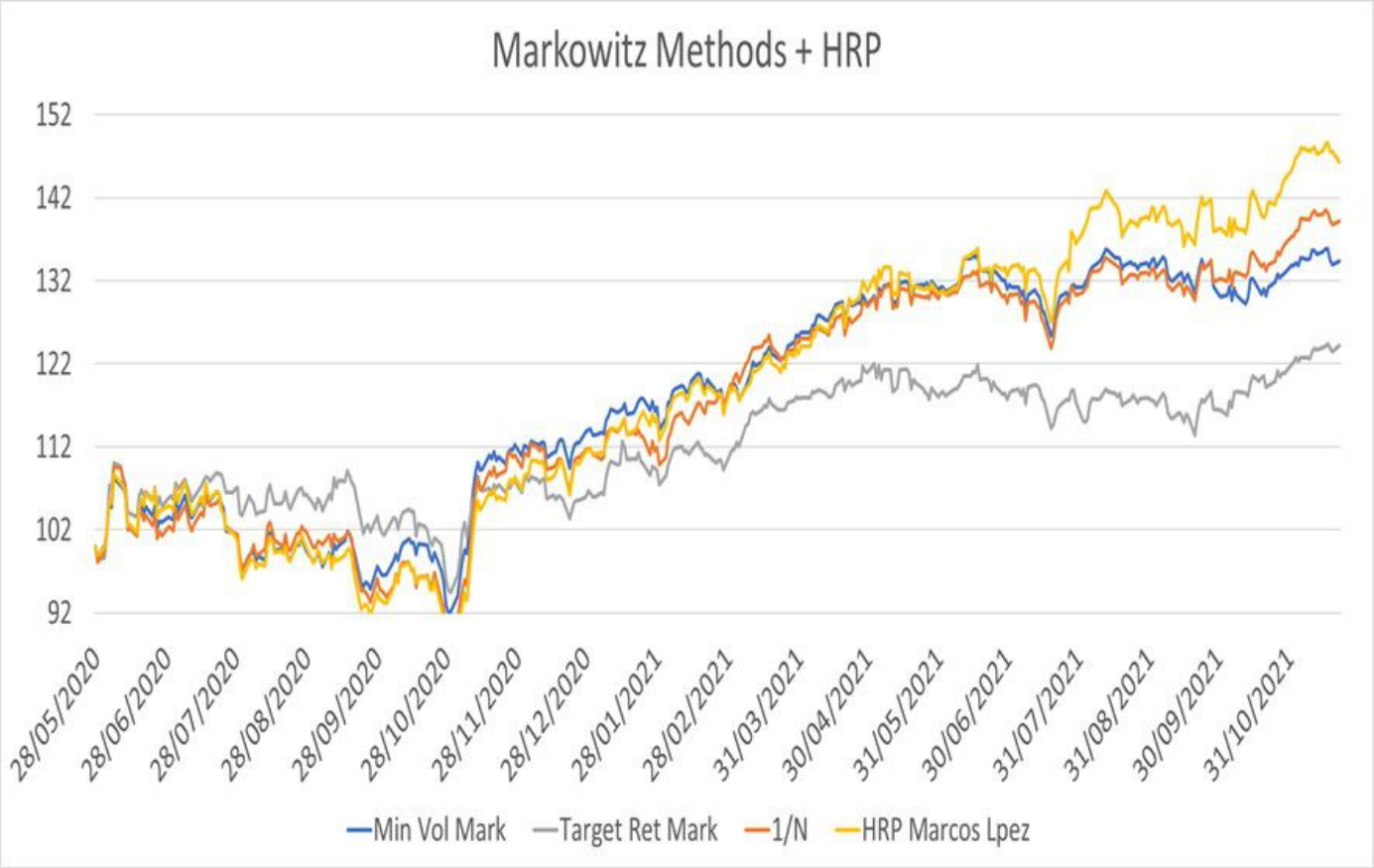}
	\caption{NAVs for EU portfolio for top mean-variance methods, 1/N and HRP: NAV starting from 01/06/2020 }
	\label{figure8}
\end{figure}

\begin{figure}
	\centering
	\includegraphics[width=130 mm]{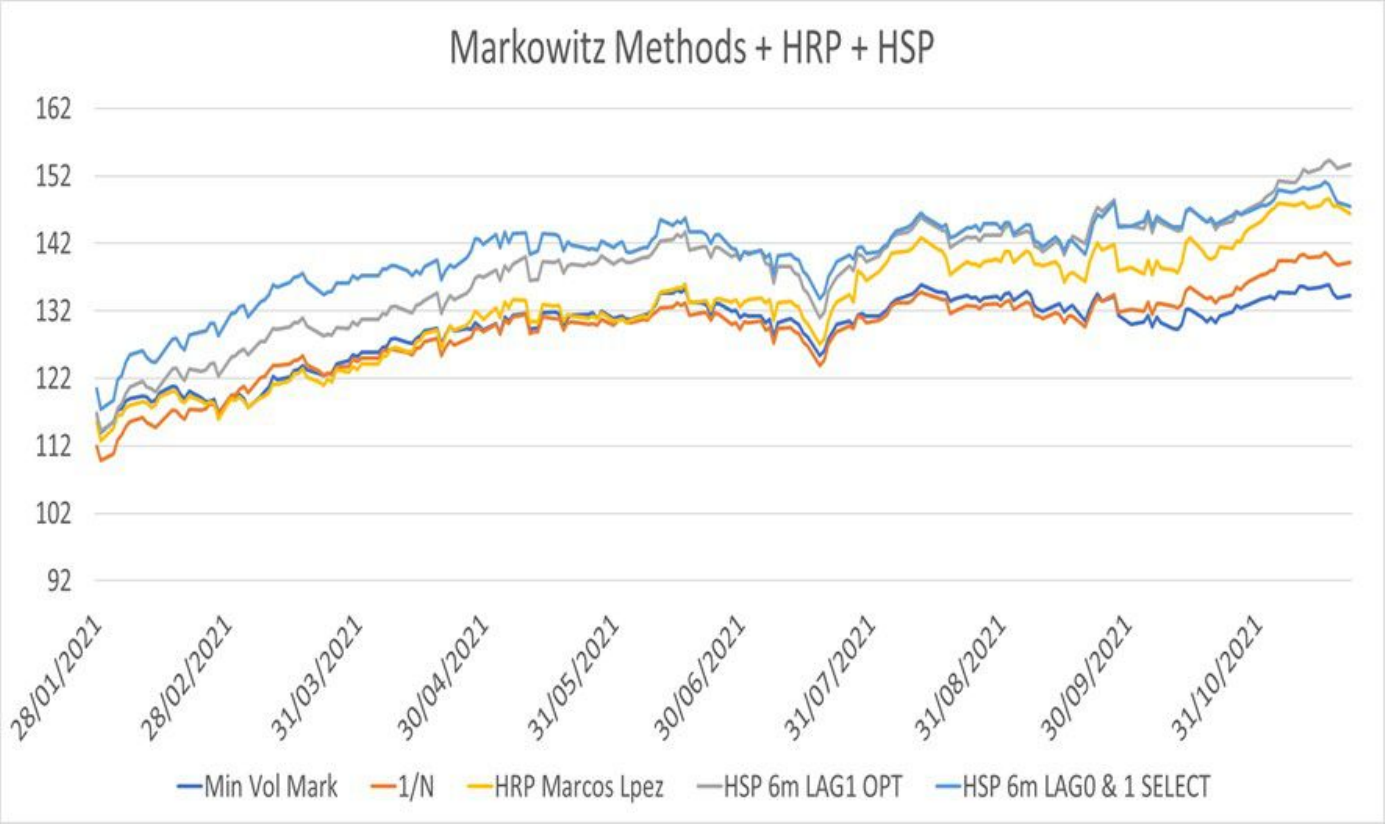}
	\caption{NAVs for EU portfolio for top mean-variance methods, 1/N, HRP, HSP for different model hyperparameters: NAV starting from 01/06/2020, showing only last sub-period}
	\label{figure9}
\end{figure}

Figure \ref{figure8} shows NAVs competitor's only, best mean-variance cases, 1/N and HRP. Figure \ref{figure9} shows the NAVs of the HSP method together with the best of the competitors. We can see how our method is better than all other methods. The same comments in the US experiment subsection are applicable to this one, and there is room for improvement in performance because we use the same configuration as in the US case. By optimizing the model hyperparameters and flexible choices we have room for better results (windows for common drivers selection, selection rationale, frequency of drivers update, fitting windows, etc.). \par 

\begin{table}[!hbt]
\centering
\begin{tabularx}{\columnwidth} {@{} l*{7}{X} @{}}
\toprule
          & Min vol  & Target Ret & 1/N    & HRP  & HSP 6m LAG 1 OPT & HSP 6m LAG0 \& 1 SELECT \\
\toprule
\midrule
Return    & 22\%         & 16\%            & 25\%   & 30\%               & 34\%             & 30\%                    \\
Vol (Ann) & 17\%         & 16\%            & 18\%   & 19\%               & 21\%             & 21\%                    \\
Sharpe    & 1,3014       & 0,9688          & 1,3740 & 1,5242             & 1,6494           & 1,433  
                \\
\bottomrule
\end{tabularx}
\caption{EU portfolio performance metrics for top mean-variance methods, 1/N, HRP, HSP for different model hyperparameters. Returns, Risk and Sharpes for full period: 01/06/2020 – 01/12/2021}
\label{table4}
\end{table}

\subsection{Long-term Investments}
\label{investments}
\begin{figure}
    \vspace*{-30mm}
	\centering
	\includegraphics[width=140 mm, height=15cm]{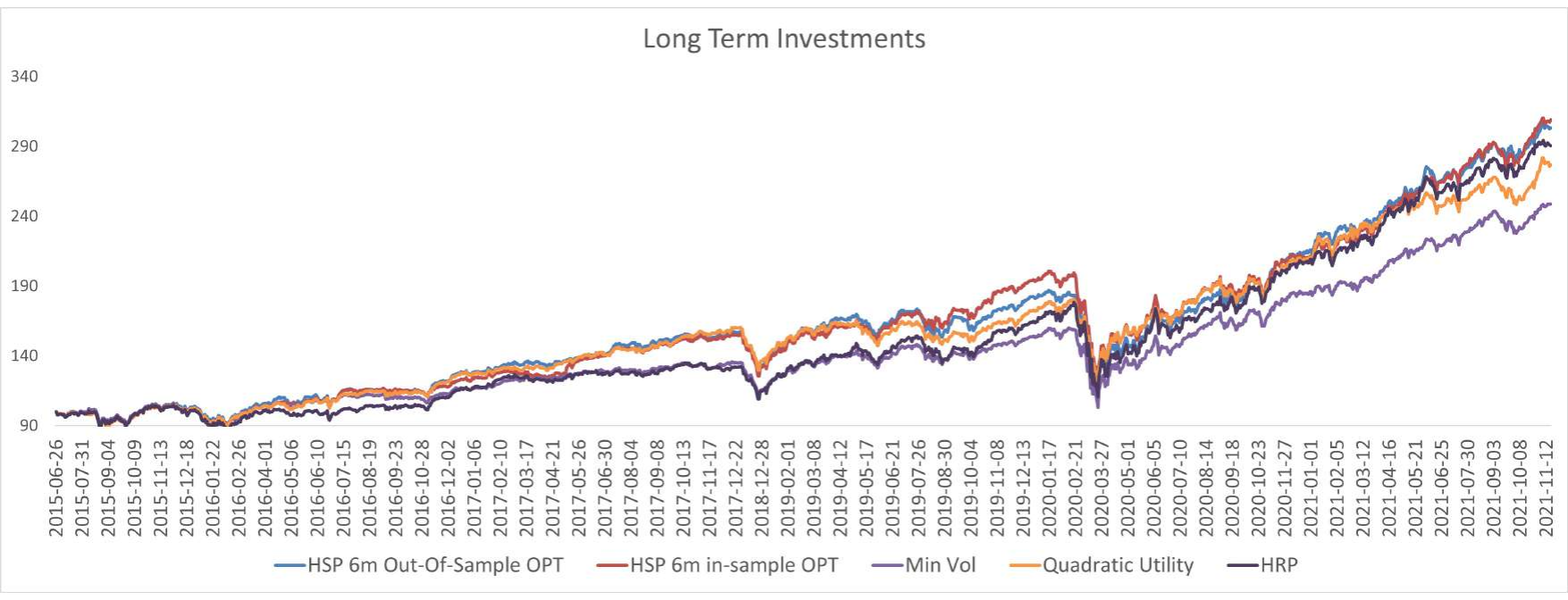}
	\vspace*{-30mm}
	\caption{NAVs for USA portfolio long-term investments for top mean-variance methods, HRP, HSP for different model hyperparameters. NAV from 06/2015 to 12/2021}
	\vspace*{-25mm}
	\label{figure17}
	\centering
	\includegraphics[width=140 mm,height=15cm]{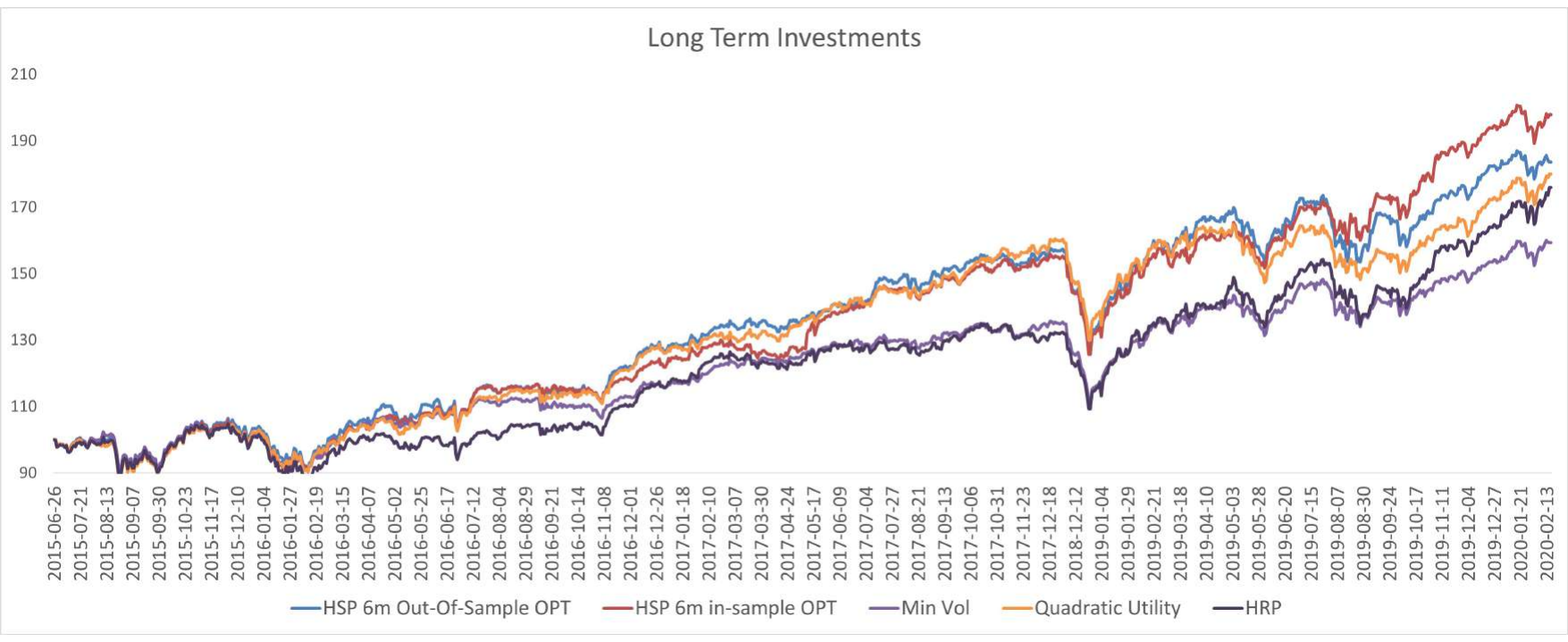}
	\vspace*{-30mm}
	\caption{NAVs for USA portfolio long-term investments zoom of Figure \ref{figure17}: from 06/2015 to pre-covid 03/2020}
	\label{figure18}
\end{figure}

We show experiments for long-term investments, for the USA portfolio from June 2015 to December 2021, with the same methodology. We see in Table \ref{table5} and Figures \ref{figure17}-\ref{figure18}, HSP outperforms all other methods (in Returns, Sharpe and NAVs). Both versions of HSP use a window of 6 months for common drivers' selection, OPT method. Out-of-sample means that the function called OptimalArchitecture(), from algorithm \ref{HSP}, uses out-of-sample test data for the averages of sensitivity values, in contrast to in-sample, which uses the averages of sensitivity values from the training data. This time common drivers' selection is updated monthly on every portfolio optimization (re-balance) date with an improvement in performance compared to the previous EU and USA cases, as seen in the NAVs series from Figure 14 and in Table 5.

\begin{table}[!hbt]
\centering
\begin{tabularx}{\columnwidth} {@{} l*{7}{X} @{}}
\toprule
          & HSP 6m Out-Of-Sample OPT  & HSP 6m in-sample OPT & Min Vol   & Quadratic Utility  & HRP
          \\
\toprule
\midrule
Return    & 18,9\%         & 19,3\%            & 15,3\%   & 17,2\%               & 18,1\%                    \\
Vol (Ann) & 21,2\%         & 21,2\%            & 19,2\%   & 20,4\%               & 21,8\%                    \\
Sharpe    & 0,89       & 0,91          & 0,80 & 0,85             & 0,83 
\\
\bottomrule
\end{tabularx}
\caption{USA portfolio long-term investments performance metrics for top mean-variance methods, HRP, HSP for different model hyperparameters. Returns, Risk and Sharpes for full period: 06/2015 – 12/2021}
\label{table5}
\end{table}

\section{Management insights}
From experiments, we find that the optimal number of common drivers depends on the number of portfolio constituents. For hyperparameter 1 between 10 and 20, we get optimal results for a portfolio of 14 constituents. With less, we are missing explanatory power and with more, multicollinearity and correlation between common drivers start being an issue. We use correlation thresholds (hyperparameters 4 and 5) so that for all drivers’ selection dates we have between 10 and 20 candidates above the thresholds. We use 6 and 12 months for correlation windows in all drivers’ selection dates (hyperparameter 2). From experiments, we find that a shorter window allows improved adaptation to changing market conditions (regimes) and works more effectively in cases where drivers’ selection is left fixed for future re-balancing dates (Subsections \ref{USAcase} and \ref{EUcase}).\par
The OPT version of the drivers’ selection following the algorithm from Section \ref{Subsection41} can use correlation thresholds, hyperparameters 4 and 5, to obtain the desired number of common drivers candidates (hyperparameter 1) for all selection dates or on average. But we can also tune hyperparameters 4 and 5, or 1, based on historical performance. The SELECT version allows the user to truncate the optimal drivers’ selection based on the algorithm in Section \ref{Subsection41} to combine common correlation following the commonality principle and other rationales. With a target hyperparameter 1, it can choose 4 and 5 to obtain a subset and fill the rest with the remaining drivers from the ranking that follow other criteria. Or for avoiding possible issues with OPT selection as already mentioned (multicollinearity, index drivers, etc.). Finally, we can truncate all the drivers’ selections and use any set of drivers we want. This is useful if we want to use smart beta indexes as drivers to use risk factor models already computed and listed as ETFs. Or use other indexes such as sector-based, or geographical indexes, credit quality indexes, or cross-asset drivers. We can use public and listed information from ETFs, funds, and indexes that are easily computed and accessible as drivers instead of hidden statistical or synthetic risk factors extracted from historical data, which can be time-consuming and computationally expensive. We perform experiments with smart beta and equities and sector indexes as common drivers that could approximate common risk factors, but we see underperformance with respect to OPT selection or a combination of OPT and SELECT possibly due to multicollinearity issues and selection bias as in \citep{doi:10.1080/00029890.2021.1965068}. A key point is that we can adapt to any frequency of updates for common drivers’ selection, to tackle back-test or model over-fitting, regimes, non-stationarity, market conditions, and changes in the portfolio at any level including risk aversion.\par
In terms of optimal network selection as in Section \ref{subsection421}, we see the optimal fitting window (hyperparameter 3) varies on each re-balancing date but is optimal from 3 to 4.5 months. For more than 4.5 months performance of the strategy goes down for the subsequent month. With respect to hyperparameter 6, we see that the optimal lag varies a lot for each re-balancing date, but it tends to work better on average if we stick to one lag for all re-balances rather than mixing both lags in a set for optimal architecture selection. In Section \ref{investments} we show two options for the user: either we use the training set for obtaining the sensitivities (in-sample) or we use the test set with different prediction horizons (out-of-sample) to obtain the sensitivity function with AAD and averages values. This is useful for users that are confident in their predictions. For the sensitivity matrix, we use Euclidean distance in the embedded space of sensitivities, but the user has flexibility in the choice of other distance metrics. For the HSP algorithm in Section \ref{HSP}, we use a single-linkage algorithm but the user has freedom in how to perform hierarchical clustering. In the experiments, we see that using a positive semi-definite neighbor of the sensitivity matrix improves performance substantially, which makes sense if we are applying recursive bisection to compute the weights \citep{Prado2016BuildingDP}.\par

\section{Conclusion}
We contribute to the state-of-the-art by incorporating sensitivity dynamics information approximated with neural networks into portfolio optimization. In contrast to factor model approaches, we use public and listed information as drivers of asset returns which are easily accessible with no computational cost and approximate the true dynamics, whereas factors can be computationally expensive to obtain, are statistical properties with many times non-realistic distributional assumptions. The sensitivity information of constituents with respect to portfolio drivers adds the directionality of portfolio returns to diversification with the dynamics of return and risk. The commonality principle selects optimal portfolio drivers in such a way that constituents can be embedded in a sensitivity space without losing idiosyncratic risk representation while adding systematic representation. We can conclude that including assets and portfolio sensitivity dynamics information with respect to public and listed drivers allows for a less computationally expensive solution and is less dependent on distributional assumptions, with true dynamics approximation adding directionality or constituents’ dynamic behavior information for diversification. We can find over-performance with respect to other out-of-sample methods and indirectly we prove over-performance with respect to risk factor methods, by using as common drivers the smart beta factors ETFs, projecting constituents into the sensitivity space of constituents with respect to these factors, and optimizing with the sensitivity matrix and HSP. \par
Secondly, we contribute by using hierarchical clustering on the sensitivity matrix to solve the convex optimization problem and incorporate hierarchical information from these sensitivities, naming it Hierarchical Sensitivity Parity (HSP). We conclude that we can improve the HRP method which is based on the hierarchies of the correlation matrix if we diversify based on the hierarchies of the sensitivity matrix. Hierarchies of the projection of constituents with respect to their common drivers add more diversification than focusing on hierarchies of projections (cosines or correlations) between assets, as seen from our experiments. Thirdly, we contribute by developing a new way to obtain maximum idiosyncratic and systematic diversification by means of the sensitivity space with respect to optimal portfolio drivers (common drivers). For this, we show that common drivers must be chosen as per the commonality principle in Section \ref{sectionGeometry}. We can conclude that systematic and maximum idiosyncratic diversification can also be reached, with the use of public and listed variables as common drivers and not just by common statistical factors allowing for extra sources of diversification such as directionality of returns from risk and return dynamics. \par
For future work, we can mention the application of other neural network models such as recurrent neural networks, or the use of neural graphs. We can also apply other distance metrics for the sensitivity matrix by exploring other geometries or statistical manifolds. We can try to explore the commonality principle in other geometries such as Bayesian networks or neural network graphs for portfolio optimization. We can investigate other ways to approximate the sensitivities or find appropriate functions for the sensitivity values apart from the average that can improve the predictive power to be included in the sensitivity matrix. Finally, we can study the use of Stochastic Differential Equations (SDEs) instead of PDEs to model asset and portfolio dynamics and approximate them with neural networks. \par

\appendix

\section{Examples of Common Drivers' Selections and Re-balancing Fittings Results for the Experiments}
\label{sec:sample:appendix}


\begin{table}[H]
\label{tablenumbers}
\centering
\begin{tabularx}{\columnwidth} {@{} l*{3}{X} @{}}
\toprule
  & 6m periods              & 1y periods              \\
\toprule
\midrule
1 & 01/12/2019 - 01/06/2020 & 01/06/2019 - 01/06/2020 \\
2 & 01/06/2020 - 01/01/2021 & 01/01/2020 - 01/01/2021 \\
3 & 01/01/2021 - 01/07/2021 & 01/07/2021 - 01/07/2021 \\
\bottomrule
\end{tabularx}
\caption{Start and end dates of past windows on the three dates for the common drivers’ selection: In the first column, we assign numbers (1,2,3) to each window for the subsequent tables. 6 and 12-month widows lengths on 01/06/2020, 01/01/2021 and 01/07/2021}
\end{table}

\begin{table}[H]
\fontsize{8}{8}\selectfont
\centering
\begin{tabularx}{\columnwidth} {@{} l*{3}{X} @{}}
\toprule
\textbf{SPX 1 6m   OPT}   & \textbf{SPX 2 6m OPT}        & \textbf{SPX 3 6m OPT}        \\
\toprule
\midrule
MSCI INDIA                & S\&P 500 HEALTH CARE IDX     & DOW JONES INDUS. AVG         \\
USD-NOK RR 25D 3M         & S\&P 500 CONS STAPLES IDX    & S\&P 500 INDEX               \\
USD-SEK RR 25D 3M         & ISHARES MSCI USA QUALITY FAC & MSCI WORLD                   \\
IBEX 35 INDEX             & ISHARES MSCI USA SIZE FACTOR & MSCI Daily TR Net World      \\
S\&P 500 HEALTH CARE IDX  & ISHARES MSCI USA MIN VOL FAC & S\&P 500 HEALTH CARE IDX     \\
S\&P 500 CONS STAPLES IDX & MSCI World Quality Pr \$     & ISHARES MSCI USA QUALITY FAC \\
STXE 600 Utilities EUR    & World Size Tilt              & ISHARES MSCI USA SIZE FACTOR \\
STXE 600 Telcomm EUR      & MSCI WORLD Min Vol PR        & ISHARES MSCI USA MIN VOL FAC \\
MSCI EM LATIN AMERICA     & MSCI World ESG               & MSCI WORLD Min Vol PR        \\
MSCI World High Dividend  & MSCI WORLD/REAL EST          & MSCI World ESG               \\
\multicolumn{2}{l}{MSCI   WORLD/HLTH CARE}               & MSCI Daily Net TR World      \\
\multicolumn{2}{l}{MSCI   WORLD/CON STPL}                &                              \\
\bottomrule
\end{tabularx}
\caption{For the three dates of common drivers' selection, for USA portfolio, and 6-month past window length, we have OPT selection of common drivers winners, which means the full algorithmic selection based on thresholds for correlation values as in  Section \ref{Subsection41} is used.}
\end{table}

\begin{table}[H]
\fontsize{8}{8}\selectfont
\centering
\begin{tabularx}{\columnwidth} {@{} l*{3}{X} @{}}
\toprule
\textbf{SPX 1 6m SELECT}          & \textbf{SPX 2 6m SELECT}           & \textbf{SPX 3 6m SELECT}\\   
\toprule
\midrule
USD SWAP SEMI 30/360 10Y & Generic 1st 'FV' Future   & DOW JONES INDUS. AVG         \\
EUR-CZK X-RATE           & BONOS Y OBLIG DEL ESTADO  & Generic 1st 'S ' Future      \\
CHF-JPY X-RATE           & NASDAQ COMPOSITE          & S\&P 500 INDUSTRIALS IDX     \\
BUONI POLIENNALI DEL TES & EUR SWAP ANN (VS 6M) 10Y  & SOYBEAN FUTURE    Nov21      \\
UK Gilts 30 Year         & U.S. TIPS                 & ISHARES MSCI USA VALUE FACTO \\
US Generic Govt 10 Yr    & MSCI World Momentum Pri\$ & ISHARES MSCI USA SIZE FACTOR \\
MSCI INDIA               & MSCI WORLD/REAL EST       & MSCI WORLD VALUE INDEX       \\
\multicolumn{2}{l}{U.S.   Treasury}                  & World Size Tilt              \\
\multicolumn{2}{l}{USD-NOK RR   25D 3M}              & MSCI WORLD/INDUSTRL          \\
\multicolumn{2}{l}{USD-SEK RR   25D 3M}              &                              \\
\multicolumn{2}{l}{NASDAQ   COMPOSITE}               &                              \\
\multicolumn{2}{l}{MSCI US REIT   INDEX}             &                              \\
\multicolumn{2}{l}{Japanese Yen   Spot}              &                              \\
\multicolumn{2}{l}{Indian Rupee   Spot}              &                              \\
U.S. TIPS                &                           &                              \\
1-3 Year EU              &                           &                              \\
\bottomrule
\end{tabularx}
\caption{Common drivers' selection with the SELECT method for final selection among the common drivers set for three different selection dates (6-months past window length). The algorithm in Section \ref{Subsection41} has been tuned based on spurious correlation, stock as part of an index driver, or multicollinearity.}
\end{table}

\bibliographystyle{elsarticle-harv}
\bibliography{cas-refs}





\end{document}